\documentclass[twocolumn,secnumarabic,amssymb,nobibnotes,aps,graphicx,pop]{revtex4-1}
\usepackage{amssymb}
\usepackage{bm}
\usepackage{graphicx}
\usepackage{amsmath}
\usepackage{color}

\def\1{{\bf 1}}
\def\0{{\bf 0}}
\def \RC {C\!_{{\scriptscriptstyle R}}}
\def \rF { F_{e{\scriptscriptstyle R}}^{{\scriptscriptstyle z}}}
\def \rU {U\!\!_{{\scriptscriptstyle R}}}
\def \rf { f_{{\scriptscriptstyle R}}}
\def \ru {u\!_{{\scriptscriptstyle R}}}
\def \rga {\gamma_e^{{\scriptscriptstyle \infty}}}

\def \E {{\cal E}}

\def \E {{\cal E}}

\def\nn{\nonumber \\}
\newcommand{\bY}{{\bm Y}}
\newcommand{\bx}{{\bm x}}
\newcommand{\bX}{{\bm X}}
\newcommand{\bu}{{\bm u}}

\newcommand{\bv}{{\bm v}}
\newcommand{\by}{{\bm y}}
\newcommand{\bz}{{\bm z}}
\newcommand{\Ba}{{\bm \alpha}}
\newcommand{\bb}{{\bm \beta}}
\newcommand{\brho}{{\bm \rho}}
\newcommand{\Be}{{\bm \epsilon}}

\newcommand{\bE}{{\bm E}}
\newcommand{\bB}{{\bm B}}
\newcommand{\bA}{{\bm A}}

\newcommand{\Bp}{{\bm p}}

\newcommand{\be}{\begin{equation}}
\newcommand{\ee}{\end{equation}}
\newcommand{\bea}{\begin{eqnarray}}
\newcommand{\eea}{\end{eqnarray}}
\newcommand{\ba}{\begin{array}}
\newcommand{\ea}{\end{array}}
\def\sq{\mbox{\rlap{$\sqcap$}$\sqcup$}}
\newenvironment{proof}[1]{\vspace{5pt}\noindent{\bf Proof #1}\hspace{6pt}}%
{\hfill\sq}
\newcommand{\bp}{\begin{proof}}
\newcommand{\ep}{\end{proof}\par\vspace{10pt}\noindent}
\begin{document}

\title{The slingshot effect: a possible new laser-driven high energy acceleration mechanism for electrons}

\author{  Gaetano Fiore$^{1,3}$, \  Renato Fedele$^{2,3}$, \
Umberto \ de Angelis$^{3}$,    \\    
$^{1}$ Dipartimento di Matematica e Applicazioni, Universit\`a di Napoli ``Federico II'',
80126 Napoli, Italy\\         
$^{2}$ Dipartimento di Fisica, Universit\`a di Napoli ``Federico II'',   80126 Napoli, Italy\\
$^{3}$         INFN, Sezione di Napoli,  80126 Napoli, Italy}


\begin{abstract}
\noindent
We show that under appropriate conditions the impact  of a very short and intense laser
pulse onto a plasma causes the expulsion of surface electrons with high energy in the
direction opposite to the one of propagation of the pulse. This is due to the combined
effects of the ponderomotive force and the huge longitudinal field arising from charge
separation  (``slingshot effect").
The effect should also be present with other states of matter, provided the
pulse is sufficiently intense to  locally cause 
complete ionization. An experimental test seems toi be feasible and,
if confirmed, would provide a new extraction and acceleration mechanism for electrons,
alternative to traditional radio-frequency-based or Laser-Wake-Field ones.
\end{abstract}

\maketitle

\section{Introduction}  

Recently developed laser technologies have allowed the production of very intense and coherent electromagnetic (EM) waves concentrated in ultra-short pulses, whose interaction with matter is characterized by fast, huge and highly nonlinear effects. 
In particular, the very strong longitudinal ponderomotive
forces caused by such laser pulses
produces very large and very fast charge separations between ions and electrons  and
thereby very strong longitudinal electric fields (in matter other than
plasma, this occurs after the huge
kinetic energy transfered to the electrons  has locally
ionized matter into a plasma). In some conditions, a plasma wave with a phase velocity
almost equal to the group velocity of the laser pulse appears behind the latter. This characterizes a mechanism known as  Laser Wake Field  (LWF) excitation \cite{Tajima-Dawson1979,Gorbunov-Kirsanov1987,Sprangle1988}, which can be used to accelerate beams of  charged particles  injected from outside (\textit{external injection} \cite{Irman2007}) or belonging to the plasma itself (\textit{self injection}  \cite{Joshi2006}).

In plasma-based acceleration schemes working with the external injection, lasers with power of up to hundreds of Terawatts are employed. Their typical intensity and pulse duration range from 10$^{20}$ to 10$^{22}$ Watt/cm$^2$ and from sub-pico to femto-seconds, respectively. In these conditions, the ponderomotive effect leads to extreme charge separation corresponding to the maximum electric field  $E_{max}$ that can be achieved in a plasma of a given unperturbed density. This is reached when the electron density perturbation becomes of the order of the equilibrium density $n_0$ and $E_{max}$[V/cm]$\approx \sqrt{n_0[cm^{-3}]}$. Then, for $n_0\sim$ 10$^{18}$ cm$^{-3}$, $E_{max}\sim$ 1 GV/cm. The ultra-strong acceleration gradients
(compared to the maximum fields that are produced in the conventional accelerating machines based
on radio-frequency cavities, which are of  the order of tens MV/m)
allow to reduce the length of the  acceleration chamber of several orders of magnitude. Then, externally injected electrons could reach an energy of the order of a 1 GeV in a few centimeters. However, the acceleration
of an externally injected beam is effective only if the initial velocity of the latter
is close to the phase velocity of the plasma wave (hence to the light speed $c$),
so that the dragging action of the wake lasts sufficiently long. Therefore,  acceleration machines
based on external injection can be used only after some pre-acceleration stage.

With a very intense laser pulse, the plasma electrons reach very high energies in very short times.
In the self injection scheme \cite{Joshi2006} the accelerated electrons are those which are naturally (albeit violently) ejected from the plasma itself. Actually, in the works of experimental feasibility, this second acceleration process, if compared to the external injection scheme, was not very efficient and reliable in terms of intensity, energy spread and collimation. Therefore, it appeared to be more suitable as an electron beam source and pre-acceleration device to be used before a subsequent acceleration stage, whether conventional or plasma-based.
However, recent investigations \cite{FauEtAl04,Kalmykov2009} have shown that acceleration schemes that are based on a violent laser-plasma interaction should work in a regime for which the laser pulse is strong enough to blow out electrons 
and leave a wake of immobile positively charged ions (\textit{bubble regime} or \textit{blowout regime}). Such a regime has been confirmed by several valuable particle-in-cell simulations. A bubble can both trap and accelerate the background electrons and it seems to be necessary to produce a very collimated, quasi-monoenergetic, self-injected electron beam. Bubbles are typically encircled by a return current that is  created by relativistic electrons. However, a reliable quantitative description is still missing. 
Remarkably, a very recent experiment with self-injection has produced quasi-monoenergetic electrons with energy well beyond 1 GeV. By applying new Petawatt laser technology, a group of researchers of Texas University at Austin \cite{Wang2013} was able to
produce electron bunches with a spectrum prominently peaked at 2 GeV with only a few percent energy spread and unprecedented sub-milliradian divergence. This circumstance seems to characterize the signature of the bubble formation that is followed immediately by a strong localized injection.

\begin{figure*}
\begin{center}
\includegraphics[height=6.3cm]{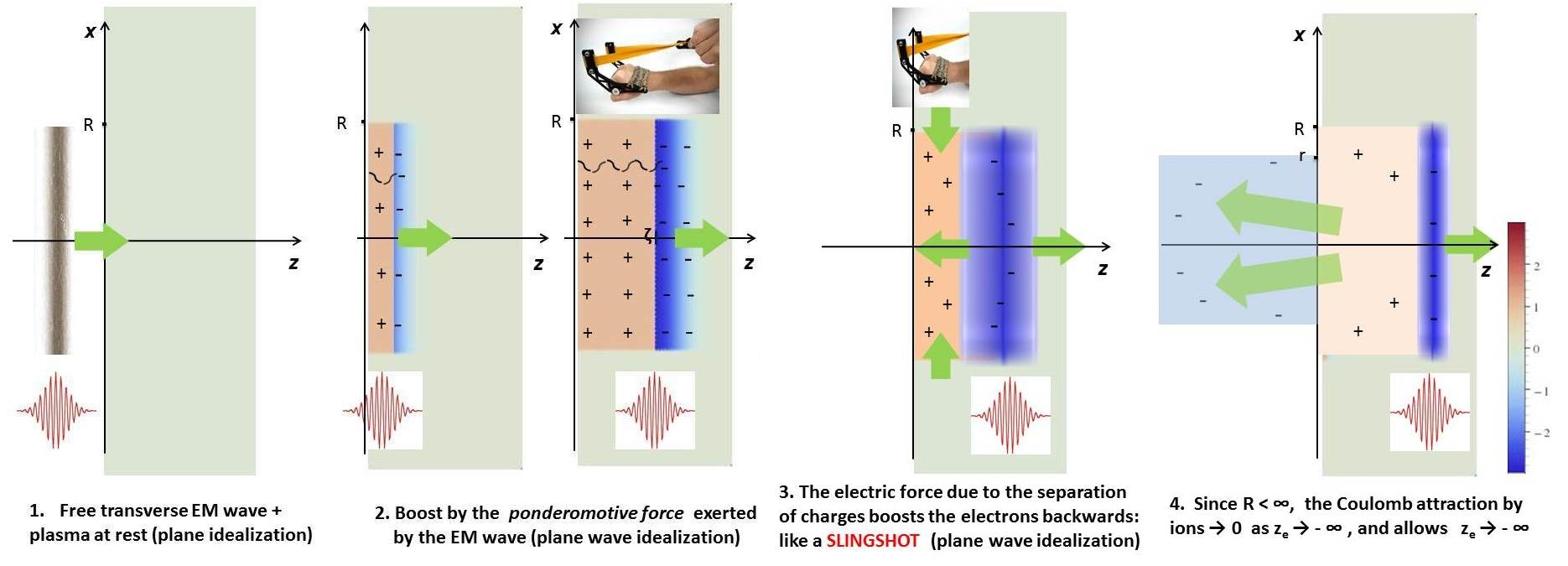} 
\end{center}
\caption{Schematic stages of the slingshot effect}
\label{plasma-laser2}      
\end{figure*}

In this paper, we show that a different acceleration mechanism of plasma electrons may occur under suitable conditions for the pulse length, duration and shape. In the self injection scheme the plasma electrons are dragged in and ejected {\it forward}, whilst in this new \textit{slingshot} mechanism \textit{surface electrons} (i.e.  plasma electrons in a thin layer just beyond the surface of the plasma) are expected to be expelled with high energy {\it backwards}. This means that, shortly after the impact of a suitable ultra-short and ultra-intense laser pulse in the form of a pancake normally onto a plasma, such electrons are expelled along the direction that is {\it opposite} to the one of the pulse propagation.  
The mechanism is very simple: the surface electrons first are given sufficient electric potential energy by displacement (through the ponderomotive force produced by the pulse) with respect to the ions, then are pulled back by the electric force exerted by the latter 
and the other electrons, and leave the plasma; provided the spot size is sufficiently small their energy will be enough to escape to infinity. The stages are schematically depicted 
in fig. \ref{plasma-laser2}. Below we show that the conditions for this to happen are that 
the pulse is sufficiently short, the radius $R$ of the pancake is sufficiently small for the EM field inside the pulse to be sufficiently intense, but also relatively large to avoid trapping of the electrons.
As this is based only on the interaction of the pulse with the first layer of plasma, a reliable,
rather explicit magnetohydrodynamic  description seems to be possible at least for a low 
density plasma. This is given here, referring the reader to \cite{Fio13} for the proof of 
some essential mathematical results (a summary can be found in \cite{Fio14}).
The results are used to suggest possible experiments
at the Italian INFN facility FLAME (Frascati Laser for Acceleration and Multidisciplinary Experiments) 
\cite{GizEtAl13}, 
and preliminary estimates for possible experiments
at more powerful facilities like the Extreme Light Infrastructure (ELI) \cite{Pri14b}.
We expect that for higher density plasmas and  suitably higher intensities of the EM field the slingshot effect still occurs and leads to the expulsion of higher energy electrons; a quantitative estimate of the latter is not possible by computations within the low density approximation considered in  \cite{Fio13}, and will be considered elsewhere.

It should be pointed out that the  interaction of very intense laser radiation with the boundary of an
{\it overdense} plasma has been studied  in \cite{Eremin2010,Geindre2010,Gonoskov2011} using particle-in-cell
 simulations or (simplified) analytical models. It was found that the effect on the EM radiation
(respectively   transmission and reflection,  absorption,   conversion from incident femtosecond
to reflected attosecond pulses) may be accompanied \cite{Eremin2010,Geindre2010} by (temporary or final)  emission of nanobunches of electrons backwards; this is the result of a  multi-cycle in-out acceleration process (described stochastically, when the hydrodynamic description fails) of the boundary electrons by a quasistationary
EM wave. On the contrary, the acceleration process presented here is induced by a
{\it short pulse} onto an {\it underdense} plasma
and produces a {\it single} bunch of electrons after a unique forward acceleration followed 
by a unique backward one; moreover, our description is purely {\it hydrodynamic}.

\section{The model}
\label{model}

We describe the plasma as a fully relativistic collisionless fluid, with the ``plasma + EM field" system fulfilling 
the Lorentz-Maxwell and the continuity equations. We assume that the plasma is initially 
neutral, unmagnetized and at rest 
with electron density equal to zero and to a constant \ $n_0$ \ respectively in the regions  \ $z\!<\! 0$ and
 $z\ge 0$. \  We consider a purely transverse EM pulse in the form of 
a pancake with cylindrical symmetry around the $z$-axis,
propagating in the positive ${\hat{\bm z}}$ direction and hitting the plasma surface at $t=0$.
Our simplified model of the impact leading to the slingshot effect consists of the following elements:

1. During the whole process we schematize the pancake pulse (see fig.
\ref{plasma-laser2}-1) as a free plane pulse multiplied by the characteristic function
of the cylinder $\RC$ of equation $\rho^2\!\equiv\!x^2\!+\!y^2\le R^2$ (with some  finite  radius $R$)
\be
\bE^{{\scriptscriptstyle\perp}}\!(t,\!\bx)=\Be\!^{{\scriptscriptstyle\perp}}\!(ct\!-\!z)\theta(R\!-\!\rho),
\quad \bB^{{\scriptscriptstyle\perp}}\!(t,\!\bx)={\hat{\bm z}}\!\wedge\!\Be\!^{{\scriptscriptstyle\perp}}\!(ct\!-\!z)
\theta(R\!-\!\rho)                                           \label{pump}
\ee
($\bE,\bB,c,\theta$ respectively stand for the electric and magnetic field, the velocity of light, 
the Heaviside step function, and we use CGS units throughout the paper);
the  `pump' function  $\Be^{{\scriptscriptstyle\perp}}(\xi)$ \ vanishes
outside some finite interval \ $0\!<\!\xi\!<\!l$. 
The pulse reaches the plasma surface \ $z\!=\!0$ at $t\!=\!0$.
We justify (\ref{pump}) at the end of  section \ref{Conclusions}.

2. We  neglect the motion of ions (this is justified at the end of  section \ref{Experiment}).

3. We first study  (section \ref{Planewavessetup}) the associated plane 
problem [$R\!=\!\infty$ in (\ref{pump})] in a parameter range allowing
 simple approximations of the forward and  backward motions of the surface electrons. 

4. We determine (section \ref{3D-effects}) a sufficiently large $R$ such that the approximation
3. is satisfactory  for the surface electrons within  some inner
cylinder $C_r$ of equation $\rho^2\!\le\! r^2$, 
with \ $r\!\ge\! R/2$.
We also estimate the corresponding deceleration of the electrons after expulsion 
and thus lower bounds for their number and final energy. 

Items $1\div 4$ and the other used approximations (see section \ref{Conclusions}) 
make the effects of the model easily computable.
Their justification consists in showing that their effect is either small
or "conservative", i.e. that it leads to {\it underestimate}  the size of the slingshot effect.
This is fine for the purpose of the present work, which is to give a safe theoretical
basis to the prediction of this new effect, rather than to optimize it.
Most proofs and detailed calculations are concentrated in the appendix.

Now some remarks on notation. We denote as \ $\bx_e(t,\bX)$  \ the position at time $t$
of the electrons' fluid element initially located at $\bX\!\equiv\!(X,Y,Z)$, and for each fixed $t$ as
$\bX_e(t,\bx)$ the inverse. 
We denote as $n_e,\bv_e$ the electrons' Eulerian
density and velocity, and shall often use the dimensionless quantities \ $\bb_e\!\equiv\!\bv_e/c$,
$\bu_e\!\equiv\!\Bp_e/mc\!=\!\bb_e/\sqrt{1\!-\!\bb_e^2}$, 
$\gamma_e\!\equiv\!1/\sqrt{1\!-\!\bb_e^2}\!=\!\sqrt{1\!+\!\bu_e^2}$. \
The Lagrangian counterparts depend on  $t,\bX$, rather than on $t,\bx$, and are denoted
 by a tilde, e.g. $\widetilde{\Bp^{{\scriptscriptstyle \perp}}}\!(t,\!\bX)=\Bp^{{\scriptscriptstyle \perp}}\![t,\!\bx_e(t,\!\bX)]$.

\subsection{Plane wave idealization}
\label{Planewavessetup}

For $R\!=\!\infty$ we choose  the gauge so that the transverse  
(with respect to ${\hat{\bm z}}$) vector potential is the physical observable 
$\bA\!^{{\scriptscriptstyle\perp}}\!(t,z)\!=\!-\!\int^{t}_{-\!\infty }\!\!dt' c\bE^{{\scriptscriptstyle\perp}}\!(t'\!,z)$;
then 
\ $\bB\!=\!\bB^{{\scriptscriptstyle\perp}}\!=\!{\hat{\bm z}}
\!\wedge\!\partial_z\bA\!^{{\scriptscriptstyle\perp}}$, \
$c\bE^{{\scriptscriptstyle\perp}}\!=\!-\partial_t\bA\!^{{\scriptscriptstyle\perp}}$.
By (\ref{pump}) it is \  $\bA\!^{{\scriptscriptstyle\perp}}\!(t,z)\!=\!
\Ba^{{\scriptscriptstyle \perp}}\!(ct\!-\!z)$, \ where
$\Ba^{{\scriptscriptstyle \perp}}\!(\xi)\!\equiv\!-\!\!\int^{\xi}_{-\infty }\!\!d\xi' \Be^{{\scriptscriptstyle\perp}}\!(\xi')$, \
implying $\Ba^{{\scriptscriptstyle \perp}}\!(\xi)=0$ for \ $\xi\!\le\! 0$.

As known, the  transverse component  of the equation of motion
$d \Bp_e/dt\!=\!-e\left(\bE \!+\!\frac{\bv}c \wedge \bB \right)$
of the electrons  implies \
$\Bp_e^{{\scriptscriptstyle \perp}} \!-\!\frac ec\bA^{{\scriptscriptstyle \perp}}\! \!\!=$const
on the trajectory of each electron; \ this is zero  at $t\!=\!0$, \  hence \
$\Bp_e^{{\scriptscriptstyle \perp}}\!=\!mc\bu_e^{{\scriptscriptstyle \perp}} \!=\!\frac ec\bA^{{\scriptscriptstyle \perp}}\!=\!\frac ec\Ba^{{\scriptscriptstyle \perp}}$.
The longitudinal component involves the electric force 
$F_e^{{\scriptscriptstyle z}}\!\equiv\!-e E^{{\scriptscriptstyle z}}$ 
and the magnetic one 
$$
\ba{l}
F^{{\scriptscriptstyle z}}_m\!\equiv\! -\frac ec(\bv_e \!\wedge\! \bB)^z \!=\!\frac{-e^2}{2m{}c^3\gamma_e}\partial_z\bA\!^{{\scriptscriptstyle\perp}}{}^2
\!=\!\frac{e^2}{2m{}c^4\gamma_e}\partial_t\Ba\!^{{\scriptscriptstyle\perp}}{}^2(ct\!-\! z). 
\ea
$$
Regarding ions as immobile, the Maxwell equations imply 
\be
E^{{\scriptscriptstyle z}}(t\!,\! z)\!=\!4\pi e n_0\!\left\{
z\,\theta(z)\!-\! Z_e\!(t\!,\! z )\, \theta[ Z_e\!(t\!,\! z )]\right\}.
\quad   \label{elF}{}
\ee
 (see e.g. eq. (46) in \cite{Fio13}). The electric force  $\widetilde{F_e^{{\scriptscriptstyle z}}}
(t,Z)\!=\!-e \widetilde{E^{{\scriptscriptstyle z}}}(t,Z)\!=\!-eE^{{\scriptscriptstyle z}}\![t,\! z_e\!(t,\!Z)]$
acting at time $t$ on the electrons initially located at $\bX$ (with $Z\!\ge\! 0$) is therefore
\bea
\widetilde{F_e^{{\scriptscriptstyle z}}}\!(t,\! Z)
\!=\! \left\{\!\!\ba{ll}
- 4\pi n_0 e^2\!  \Delta z_e\:\:  \mbox{(elastic force)}\: &\mbox{if }z_e\!>\!0,\\[4pt]
\:\: 4\pi n_0 e^2 Z\:\:  \mbox{(constant force)}\:\: &\mbox{if }z_e\!\le\!0,
\ea\right.\quad           \label{conservativeF}
\eea
where $\Delta z_e\!\equiv\! z_e(t\!, Z)\!-\! Z$ is the electron displacement
with respect to its initial equilibrium position $Z$. 
$\widetilde{F_e^{{\scriptscriptstyle z}}}$ is thus the force 
 of a harmonic oscillator (with equilibrium at $z_e\!=\!Z$)
in the bulk and a nonnegative constant outside.

Assume that $\Be^{{\scriptscriptstyle \perp}}\!=\!\epsilon_s\Be_o^{{\scriptscriptstyle \perp}}$, where
$\epsilon_s(\xi)\!\ge\! 0$ slowly varies inside the interval \ $0\!<\!\xi\!<\!l$
and is zero outside, \ while $\Be_o^{{\scriptscriptstyle \perp}}(\xi)$ 
sinusoidally oscillates around zero
with a period $\lambda\!\ll\! l$. If the pump $\Be^{{\scriptscriptstyle\perp}}$ is very large, by continuity
we expect the EM field to remain close to the compact-support travelling-wave
(\ref{pump})  also for small $t\!>\!0$. 
In Ref. \cite{Fio13} it is shown that this is indeed the case in the space-time region  
$0\!\le\!ct\!-\!z\!\le\! \xi_0$, $0\!\le\!ct\!+\!z\!\ll\!  \frac{2\pi} {K\lambda}$, \ where $\xi_0$ is the first maximum point of $\epsilon_s(\xi)$, and \ $K\!\equiv\!\pi e^2 n_0/m{} c^2$. \

The first effect of the impact of the pulse is to boost {\it all} the electrons reached not only in  
the $x,y$ directions through the electric force  
$-e\bE^{{\scriptscriptstyle \perp}}\!=\!-e\Be^{{\scriptscriptstyle\perp}}$, but also
in  the   $z$  direction through $F^{{\scriptscriptstyle z}}_m$. Since
$\Bp_e^{{\scriptscriptstyle \perp}}\!=\!\frac ec\Ba^{{\scriptscriptstyle \perp}}$ oscillates
about zero for $0\!\le\!\xi\!\le\! l$, and $\Ba^{{\scriptscriptstyle \perp}}\!(\xi)\!=\!\mbox{const}\!\simeq\!0$ 
for \ $\xi\!\ge\! l$, then the transverse  motion approximately averages to zero, and
 $\Bp_{e}^{{\scriptscriptstyle \perp}}\!\simeq\!0$  after the pulse.
Since $\Ba\!^{{\scriptscriptstyle\perp}}{}^2$ is zero at the front
of the pulse and positive inside,  its time derivative is positive at the front;
hence the initial $z$-boost is necessarily  {\it positive}.
Because of this boost, the first layer of ions remains unshielded while electrons accumulate just beyond
the surface $S_0(t)$ of discontinuity of $n_e$ delimiting the $Z\!=\!0$ surface electrons 
[the resulting charge distribution in the   approximation (\ref{completeelectronmotion})  to be considered here
is depicted in the first  four upper pictures of fig. \ref{ChargeDistributions}].
This charge separation generates the {\it slingshot}, {\it i.e.} the longitudinal electric force
$\widetilde{F_e^{{\scriptscriptstyle z}}}$ \ 
pushing the boosted electrons backwards. 
If the time scale  of significant variations of $\epsilon_s$ (and the duration $l/c$ of the pulse, {\it a fortiori})
is much larger than the characteristic period $\tau\!\gtrsim\!T\!_{{\scriptscriptstyle H}}^{{\scriptscriptstyle nr}}$
of oscillation of the electrons ($T\!_{{\scriptscriptstyle H}}^{{\scriptscriptstyle nr}}\!\equiv\!\sqrt{\pi m/n_0e^2}\!=\!2\pi/\omega_p$ 
is the period of the corresponding nonrelativistic harmonic oscillator)
\cite{footnote1},
then the electrons  do many oscillations during the pulse, 
the power $P\!\equiv \!\widetilde{F_{{\scriptscriptstyle m}}^{{\scriptscriptstyle z}}}
\widetilde{v^{{\scriptscriptstyle z}}}$ transfered  by the magnetic force to each electron
oscillates about zero, and its time integral $\E_p$ (the transfered energy, or {\it slingshot loading}) is approximately zero. 
This was the situation normally encountered in laboratories until a few years ago, because of the inability to generate sufficiently low densities and short pulses.
If $l/c\!<\!T\!_{{\scriptscriptstyle H}}^{{\scriptscriptstyle nr}}/4$, then $\widetilde{v^{{\scriptscriptstyle z}}}$ 
is positive during
all the pulse while $\widetilde{F_{{\scriptscriptstyle m}}^{{\scriptscriptstyle z}}}$ oscillates
about zero, and overall we obtain only a moderate slingshot loading $\E_p\!>\!0$ 
(if $l/c\!\ll\!T\!_{{\scriptscriptstyle H}}^{{\scriptscriptstyle nr}}$ a good approximation of the motion is the zero-density solution recalled below). 
Therefore to increase $\E_p$ we study the range $l/c\!\sim\!T\!_{{\scriptscriptstyle H}}^{{\scriptscriptstyle nr}}$.
Denoting as $\langle \: \rangle$  the average over a period $\lambda$, we find \ $F_{{\scriptscriptstyle p}}^{{\scriptscriptstyle z}}\equiv\langle F^z_m \rangle \propto\partial_t \epsilon_s^2(ct\!-\!z)$, \
and the ponderomotive force \ $F_{{\scriptscriptstyle p}}^{{\scriptscriptstyle z}}$
is respectively positive, negative when $\epsilon_s^2$ is strictly increasing, decreasing - as known.
If for simplicity $\epsilon_s^2(\xi)$ has a unique maximum point $\xi_0$, 
to maximize $\E_p$ we should make $\widetilde{v^{{\scriptscriptstyle z}}}$ switch from positive
to negative only once during the pulse, exactly at  the time $\bar t$ when the maximum  reaches the  electrons,  so that
$\langle P\rangle\!\simeq\! F_{{\scriptscriptstyle p}}^{{\scriptscriptstyle z}}\widetilde{v^{{\scriptscriptstyle z}}}$ 
keeps  nonnegative during all the motion. 
This can be  achieved by tuning $n_0$ and the  pulse length in the range
\be
l\sim cT\!_{{\scriptscriptstyle H}}^{{\scriptscriptstyle nr}}/2=\sqrt{\pi mc^2/4n_0e^2},  \label{cond1}
\ee
as well as the shape of the pulse \cite{footnote2}.
After overcoming the surface electrons, the EM pulse will further propagate forward, slowly damped;
it may also generate a wake with deeper electrons  in the plasma.
We shall ignore such phenomena and follow only the motion of the surface  electrons,
showing that a thin layer is finally expelled with high energy in the backward direction. 

The above picture is confirmed by the first steps of an iterative resolution scheme 
of the plane problem proposed in  \cite{Fio13}. The initial step is  the `zero-density' 
solution, which we summarize in eq. (\ref{n=0'+}-\ref{hatdtzz'}). 
According to it, the longitudinal motion of any electron is never backward. The $Z\!=\!0$ electrons are
reached by the maximum of $\epsilon_s^2$ at time $t_0\!=\!\Xi_e(\xi_0)/c$, 
where $\Xi_e$ is defined in (\ref{defYXi}). The error with respect to the real solution increases with $t$.
In the second step (first correction  to the `zero-density' motion) these electrons invert
their motion at a time $\bar t_1$ given in subsection \ref{Planesetup}. Electrons initially located at small $Z\!>\!0$ 
will approximately move in the same way, after the space-time displacement $(Z,Z/c)$.
For technical simplicity in this work we stick to parameters in a range [see conditions
(\ref{cond}) and the paragraph of eq. (\ref{Kxi1})]
such that  the relative difference between the zero-density and the first corrected motion keeps much smaller
than 1 for $t\!\le\! t_0$ and rapidly grows for $t\!>\! t_0$, so that $\bar t_1\!-\!t_0$ is positive, but as small
as possible. This ensures \cite{Fio13}  that: i)  the `zero-density' motion (\ref{n=0'+}-\ref{hatxtxp}) 
is a good approximation of the real forward motion for all electrons
initially located in a thin superficial layer, 
in particular it can be used to estimate their
maximal displacement $\zeta$:
\be
\zeta \,\simeq\,Y_e^z(\xi_0),\qquad Y_e^z\!(\xi)\equiv\frac{e^2}{2m{}^2c^4}\!\!\int_0^\xi\!\!\!\! 
dy\:\Ba\!^{{\scriptscriptstyle \perp}2}(y);
\label{zeta}
\ee
ii) in this approximation the slingshot loading is efficient.

\begin{figure*}
\includegraphics[width=16cm]{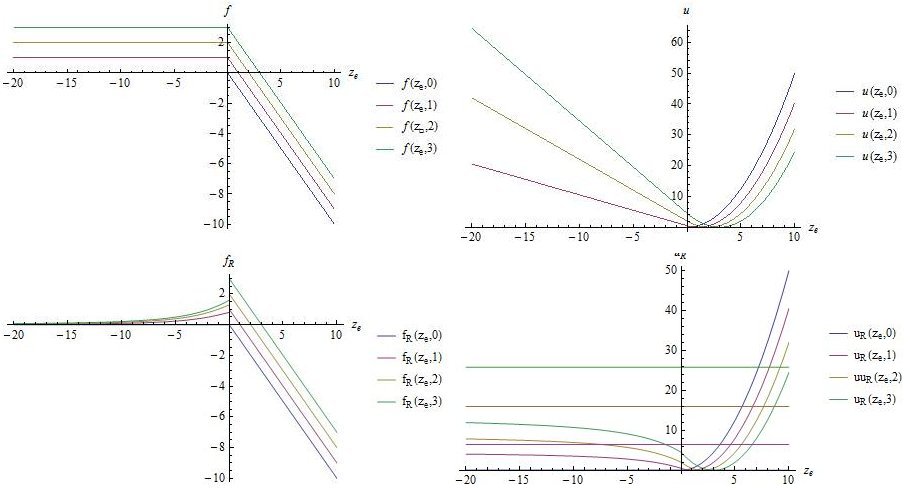}
\caption{The rescaled longitudinal electric force $f$ (left, up) and the associated rescaled potential energy
$u$ (left, down) in the idealized plane wave case, the  rescaled longitudinal electric force $\rf$ (right, up)
and the associated rescaled potential energy $\ru$ (right, down) in the case of a pancake of radius $R=5$,
plotted as functions of $z_e$ for
$Z=0,1,2,3$; the horizontal lines in the right down graph are the left asymptotes of $u$ for
$Z=0,1,2,3$.}
\label{PlotElForce}
\end{figure*}

In the backward motion of the surface electrons for simplicity we underestimate 
as \ $\widetilde{F_e^{{\scriptscriptstyle z}}}(t,Z)$ \ the 
longitudinal force, neglecting the backward $\widetilde{F_m^{{\scriptscriptstyle z}}}$
due to the last part of the pulse  and of the generated `reflected' EM wave (see the final section).
We denote the potential energy associated to the conservative force \
$F_e^{{\scriptscriptstyle z}}(z_e,Z)\!\equiv\!4\pi e^2 n_0\left[Z\!-\! z_e\,\theta(z_e)\right]$ \ 
of (\ref{conservativeF})  as \cite{footnote3}
\be
U(z_e,Z)=2\pi n_0e^2[ \theta(z_e)z_e^2\!-\!2z_eZ\!+\!Z^2].   \label{elPE}{}
\ee
In fig.  \ref{PlotElForce}-left we plot $f\!\equiv\!F_e^{{\scriptscriptstyle z}}/4\pi e^2 n_0$
and $u\!\equiv\!U/2\pi e^2 n_0$ as functions of $z_e$ for a few values of $Z\!\ge\!0$.
Using energy  conservation during the backward motion 
\bea
H\!\equiv\!
m{}c^2\gamma_e(z_e,\!Z)\!+\!U(z_e,\!Z)  \!=\!
m{}c^2\!\sqrt{\!1\!+\!\bu_{ei}^{{\scriptscriptstyle \perp}2}}
\!+\!2\pi n_0e^2\zeta^2\!\nn 
 \!\ge\!  m{}c^2\!+\!2\pi n_0e^2\zeta^2, \qquad\quad \:
\label{conservation}
\eea
one can compute $\gamma_e\!\equiv\!1/\sqrt{1\!-\!\bv_e^2/c^2}$, and therefore also
$\beta^z_e$,  as functions of
$z_e,Z$ (here $mc\bu_{ei}^{{\scriptscriptstyle \perp}}$ is the electron transverse momentum 
when the displacement is maximal, $z_e\!-\!Z\!=\!\zeta$), and the longitudinal
motion by quadrature (appendix \ref{Planesetup});
the combined forward-backward motion is summarized 
in eq. (\ref{completeelectronmotion}) and plot in  fig. \ref{zePlot}  for some fixed, 
relevant values of the parameters and a few different values of $Z$ (see below); 
as the corresponding map $Z\mapsto z$ is one-to-one for all
$t$, different worldlines do not intersect. This shows that in this
time lapse our treatment of the electron fluid as collisionless is consistent.
The mechanical energy of the $Z\!=\!0$  electrons after
expulsion ($z_e\!<\!0$)  is purely kinetic, because $U(z_e,\!0)\!=\!0$. 
However, as $U(-\infty,\!Z)\!=\!\infty$ for $Z\!>\!0$, one concludes that
in the $R\!=\!\infty$ idealization only the former electrons escape to
$z_e\!=\!-\infty$ infinity; inner electrons  invert their motion
where $\gamma_e$ reaches its minimum $\simeq\! 1$ 
and then oscillate around $Z$.  We now show that  $R\!<\!\infty$ allows the escape
 of a thin layer  of electrons.

\subsection{Inclusion of 3D-effects}
\label{3D-effects}

In the realistic case of a finite  radius $R$ the ponderomotive force of the pulse will boost only the electrons
approximately  in the cylinder $\RC$ of the same radius. 
We require $R$ to be sufficiently large with respect to $\zeta$ in order that: 
\ i) \ the electrons in a cylinder $C_r$ 
of radius $r\!>\! 0$ undergo approximately the same motion (\ref{completeelectronmotion}) as with $R\!=\!\infty$;
 \ ii) \ there is no trapping, i.e. the way backwards out for the surface electrons within $C_r$ is not obstructed by the
electrons initially located just outside the lateral surface $\partial \RC$ of $\RC$  (the latter electrons
first are boosted outward, because on  $\partial \RC$ so is directed the gradient of \
$\bE^{{\scriptscriptstyle\perp}2}$ of the pulse,
then move towards the $z$-axis  attracted by the ions). 
At the end of appendix \ref{outside}  we show that both requirements are satisfied by the condition
\be
R\gtrsim \sigma \zeta,          \label{nobubble}
\ee
with $\sigma$  a little larger than 1 (at least in the range of relevant conditions). 
We now show that: \ iii) \ due to the finite $R$,  outside the bulk
the attracting force by the ions decreases sufficiently fast
with $|z_e|$ to allow a thin layer of the expelled electrons
to escape to infinity. 

\begin{figure}
\includegraphics[width=8cm]{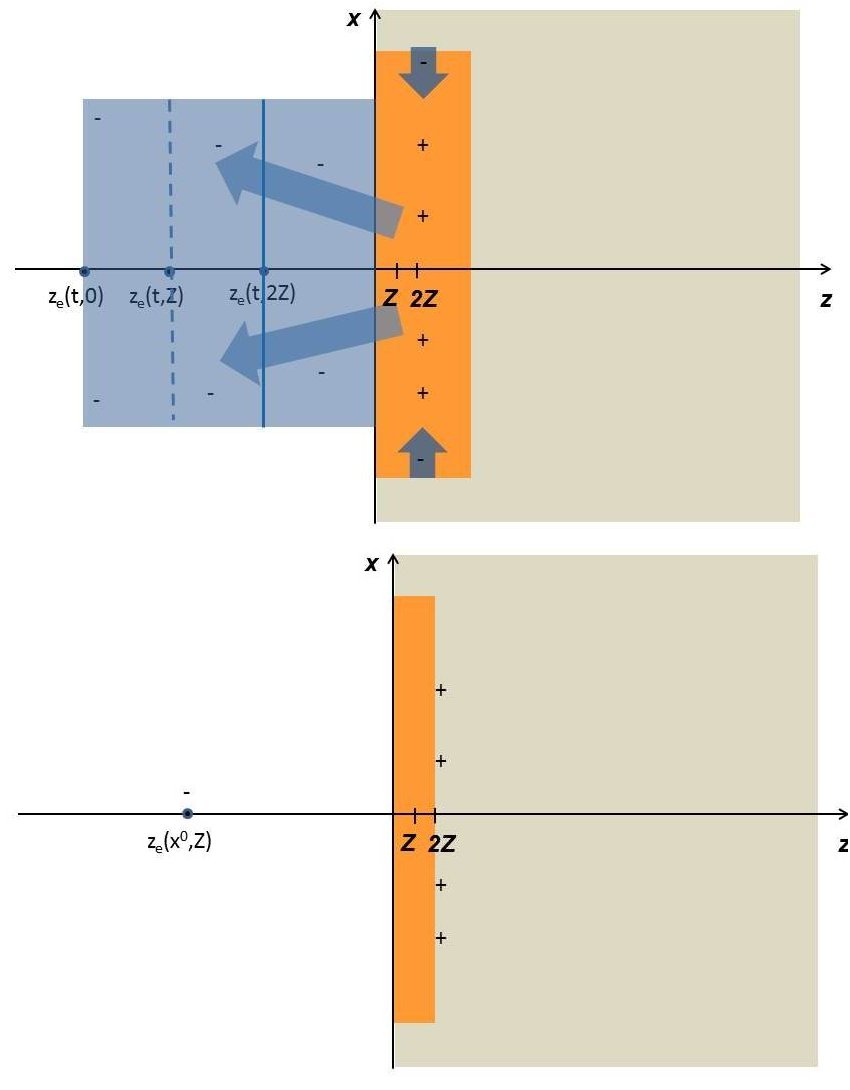}
\caption{Up: schematic picture of the expected charge
distribution shortly after the expulsion  (long arrows) of surface electrons; short arrows 
represent the inward motion of lateral electrons. 
Down: simplified charge distribution.
}
\label{pancake}
\end{figure}

\begin{figure}
\includegraphics[width=8cm]{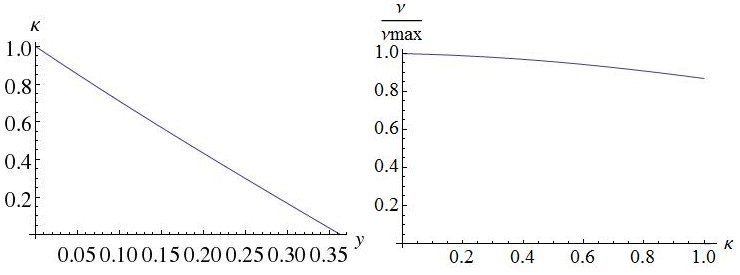}
\caption{$\kappa(y)$ vs. $y\in [0,y_{{\scriptscriptstyle M}}]$ \
($y_{{\scriptscriptstyle M}}\!\equiv\!Z_{{\scriptscriptstyle M}}/\zeta$)
(left), \ and the
kinetic energy spectrum $\nu(\kappa)$ [in units of $\nu_{max}\!=\!\nu(\kappa\!=\!1)$] \ vs.
$\kappa\!\in\! [0,1]$ \  (right), \ for $R\!=\!1.5\zeta$.
}
\label{gammaplot}
\end{figure}

As said $\widetilde{\Bp_{e}^{{\scriptscriptstyle \perp}}}\!\simeq\!0$ after the pulse.
We consider only the motion of the electrons
moving along the $\vec{z}$-axis ($\rho\!=\!0$) after the pulse; in other words we consider
those electrons which experience the strongest restoring force after expulsion,
by symmetry reasons. In fig. \ref{pancake}-up we depict the expected charge
distribution  of the electrons initially located at $Z\!\gtrsim\! 0$ \ at a time $t$ shortly after their expulsion.
The light blue area is occupied only by electrons. The left border, the dashed line and the solid line
respectively represent the surfaces $S_0,S_1,S_2$ occupied by
the electrons initially located at the points $\bX'\!\in\! C_r$ such that
$Z'\!=\!0,Z,2Z$. The orange area is positively charged due to an excess of ions.
We can bound the {\it real} electric longitudinal force
$\widetilde{F_e^{{\scriptscriptstyle zr}}}$ experienced by the electrons moving along
the  $\vec{z}$-axis as follows:
\be
0\le \widetilde{F_e^{{\scriptscriptstyle zr}}}\!(t,\!Z)
=-e\widetilde{E_{{\scriptscriptstyle -}}^{{\scriptscriptstyle z}}}\!(t,\!Z)\!-\!e
\widetilde{E_{{\scriptscriptstyle +}}^{{\scriptscriptstyle z}}}\!(t,\!Z)\le \rF\![z_e(t,\!Z),\!Z].
\ee
Here $\widetilde{E_{{\scriptscriptstyle -}}^{{\scriptscriptstyle z}}}(t,Z)$ stands for the part of the
longitudinal electric field
generated by the electrons between $S_0,S_2$;
since those between $S_0,S_1$
have by construction the same charge as those between $S_1,S_2$, but are more dispersed, it will be
$-e\widetilde{E_{{\scriptscriptstyle -}}^{{\scriptscriptstyle z}}}(t,Z)\le 0$. The part
$-e\widetilde{E_+^{{\scriptscriptstyle z}}}(t,Z)$  of the longitudinal electric
force generated by the ions and  the remaining electrons (at the right of $S_2$)
 will be smaller than the force $\rF$ generated
by the charge distribution depicted in fig. \ref{pancake}-down, where  the remaining
electrons are located  farther from $(0,0,z_e)$ (not in their actual positions, 
but in their initial ones $\bX'$) and therefore generate a smaller repulsive force.
Using cylindrical coordinates $(Z'\!,\rho'\!,\varphi')$ for $\bX'$ one easily finds  for $z_e\!\le\!0$
\bea
&& \rF(z_e,Z)
\equiv\int_0^{2Z}\!\!\!dZ'\!\!\! \int_0^R\!\!d\rho' \,
\frac {2\pi n_0e^2\rho'(Z'\!-\!z_e)}{\big[\rho'{}^2\!+\!\left(Z'\!-\!z_e\right)^2\big]^{3/2}}\!\nn && 
=2\pi n_0e^2 \!\!\left[2Z\!+\!\sqrt{\! z_e^2\!+\!R^2}\!-\!\sqrt{\! (2Z\!-\!z_e)^2\!+\!R^2}\right].\nonumber
\eea
$\rF$ is nonnegative and goes to zero as $z_e\!\to\!-\infty$, as it must be; it
 becomes a function of $t$ through $z_e(t,Z)$ only.
The associated potential energy  is \cite{footnote4}
$$
\ba{lll}
\rU(z_e,\!Z) \!\!&\!\!=\!\!&\!\! \pi n_0e^2 \!\left[(z_e\!-\!2Z)\sqrt{\! (z_e\!-\!2Z)^2\!+\!R^2}\!-\!4Zz_e +\right. \\[6pt]
&&R^2\sinh^{-1}\!\!\frac{z_e\!-\!2Z}R\!-\!z_e\sqrt{\! z_e^2\!+\!R^2}
 \!-\!R^2\sinh^{-1}\!\!\frac{z_e}R  \\[6pt]
&& \left.\!+ 2Z^2\!+\!2Z\sqrt{\! 4Z^2\!+\!R^2} \!+\!R^2\sinh^{-1}\!\!\frac{2Z}R\right],
\ea
$$
a decreasing function of $z_e$ with finite left 
asymptotes 
for any $Z\!\ge\! 0$.
In fig. \ref{PlotElForce}-right  the plots of \
$\rf\!\equiv\!\rF/4\pi n_0 e^2$,
\ $\ru\!\equiv\!\rU/2\pi n_0 e^2$   replace  those of $f,\! u$ 
 for $z_e\!\le\! 0$.
Denote as
\bea
&&\!\!\!\!\!\!\rga\!(Z) =1\!+\![2\pi n_0e^2\zeta^2\!-\!\rU(-\infty,\!Z)]/m{}c^2 \label{maxgamma} \\[4pt]
&&\ba{l}
=1\!+\! K\left[2\zeta^2\!+\!2Z^2\!\!-\! 2Z\sqrt{\! 4Z^2\!+\!R^2}
\!-\!R^2\sinh^{-1}\!\!\frac{2Z}R\right]\!          \nonumber
\ea
\eea
the $z_e\!\to\!-\infty$ limit of the relativistic factor $\gamma_e$ determined by
replacing $U$ by $\rU$ with  $\bu_{ei}^{{\scriptscriptstyle \perp}2}\!=\!0$ in  (\ref{conservation}),
and as $Z_{{\scriptscriptstyle M}}$ the value of $Z$ for which $\rga(Z)\!=\!1$.
As said, by such a replacement we overestimate the restoring rightwards force
experienced by the electrons for $z_e\!<\!0$. Consequently, the real relativistic factor
of the electrons will be larger than the above $\gamma_e$, and the
$M_0\!\equiv\!\pi r^2Z_{{\scriptscriptstyle M}}n_0$
electrons in $C_r$ with \  $0\!\le\! Z\!\le\!Z_{{\scriptscriptstyle M}}$ \ 
will be only part of those escaping to infinity;
a lower bound for their final relativistic factor is  \ $\rga(Z)$. \
We find also the  bound $|Q|\!\ge\!e\pi n_0 r^2Z\!_{{\scriptscriptstyle M}}$ 
for the total expelled electric charge $Q$.
In fig. \ref{gammaplot}-left we plot  the normalized kinetic energy \
$\kappa\!\equiv\![\rga\!-\!1]/2K\zeta^2$
 \ as a function of  \ $y\!\equiv\!Z/\zeta$ \    for
$R\!=\!1.5\,\zeta$ (by definition
$0\!\le\!\kappa \!\le\!1$, $0\!\le\! y\!\le\!y_{{\scriptscriptstyle M}}$, where
$y_{{\scriptscriptstyle M}}\!\equiv\!Z_{{\scriptscriptstyle M}}/\zeta$). 
The plot for $R\!=\!1.25\zeta$ does not differ significantly.
The fraction of expelled electrons with initial position
$Z'\in[Z,Z\!+\!dZ]$ is no less than $\pi r^2n_0dZ=\pi r^2\zeta n_0dy$ and, using $\rga$ instead of
the real final relativistic factor, the fraction with kinetic energy
in the interval $[\kappa,\kappa\!+\!d\kappa]$
is no less than $\nu(\kappa)d\kappa$, where  
\be
\nu(\kappa)\!=\!-\frac{\pi r^2\zeta n_0}{\kappa'(y)|_{y=\hat y(\kappa)}}\!=\!\frac{\pi r^2\zeta n_0}
{2\big[\sqrt{\!4y^2\!+\!R^2/\zeta^2}\!-\!y\big]_{y=\hat y(\kappa)}}
\ee  
 [$\hat y(\kappa)$ is the inverse of  $\kappa(y)$] represents the
associated energy spectrum; this is plotted in fig. \ref{gammaplot} right.
By (\ref{maxgamma}) the final relativistic factor of the expelled electrons  is
\be
\gamma_{e{\scriptscriptstyle M}}\equiv\rga(0)\simeq  1+2K \zeta^2\simeq
1+2K \,\left[ Y^z_e\!(\xi_0)\right]^2;       \label{gammaeM}
\ee
the last $\simeq$ holds under condition (\ref{cond}).

We expect $\zeta,\gamma_{e{\scriptscriptstyle M}}$ to grow with the intensity
- although at a slower rate - even if  (\ref{cond}) is not fulfilled.

\section{Experimental feasibility} 
\label{Experiment}

\begin{figure}
\includegraphics[width=7.8cm]{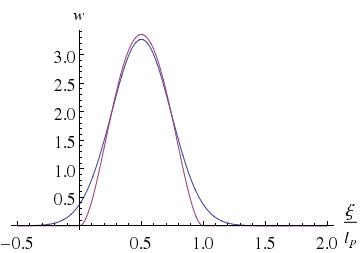}
\caption{Graphs of Gaussian (blue curve) and corresponding cut-off polynomial (\ref{modamp}) (purple curve)
normalized modulating amplitude  $w(\xi)\!\equiv\!\frac{ e\lambda}{2\pi mc^2}\epsilon_s(\xi)$.}
\label{modAmp}       
\end{figure}

We now briefly discuss the experimental conditions for the slingshot effect,
in particular for possible experiments at the FLAME facility in Frascati (but several other 
laboratories have comparable lasers) and at future ELI infrastructures.
Laser pulses of wavelength $\lambda$,
length $l\!\gg\!\lambda$, for simplicity symmetric around $\xi_0=l/2$, concentrated onto an area $\pi R^2$,
have an energy
\be
\E=\!\int\!\!dV\frac{\bE^{{\scriptscriptstyle\perp}2}\!\!+\!\bB^{{\scriptscriptstyle\perp}2}}{8\pi}
\!\simeq\!\frac{(\pi R)^2}{\lambda^2}\!\! \int_0^l\!\!\!\!d\xi \bA\!^{{\scriptscriptstyle\perp}2}
\!=\!\frac{4(\pi mc^2\! R)^2}{(e\lambda)^2} Y_e^z\!(\xi_0).           \label{pulseEn}
\ee
(we have used the relation $\bE^{\scriptscriptstyle\perp}\!\simeq\! \bA^{\scriptscriptstyle\perp}2\pi/\lambda$,
valid for a  modulated approximately monochromatic plane wave).
By (\ref{zeta}) \
$\zeta\!\simeq\!\E (e\lambda)^2 /4(m{}\pi c^2R)^2$, \ and (\ref{nobubble}) takes the form
$$
\ba{l}
R^3\gtrsim 
\frac{\sigma\E (e\lambda)^2}{4(\pi m{}c^2)^2}
\ea
$$
To maximize $\zeta$ and $\gamma_{e{\scriptscriptstyle M}}$
we choose $R\!=\!\sigma\zeta$; we obtain
\be{}
\ba{l}
\sigma\zeta\!=\!R\!=\!\left[\!\frac{\sigma\E (e\lambda)^2}{4(\pi m{}c^2)^2}\!\right]^{\frac 13}\!\!\!.\qquad
\ea    
        \label{blabla}
\ee
The  laser at the FLAME facility can
shoot  linearly polarized pulses with  \ $\lambda\!\simeq\! 8\!\times\! 10^{-5}$cm, \
energy \ $\E\!=\!5 \!\times\! 10^{7}$erg, \  and
an approximately gaussian modulating amplitude
 with width at half height \ $l'\!\simeq\! 7.5\!\times\! 10^{-4}$cm \
\cite{JovFedTanDeNGiz12,GizEtAl13}.
In the appendix we show that it is sufficient to choose  $\sigma\!\equiv\! R/\zeta\!=\!1.25,\, 1.5$ \ to  obtain 
the expulsion of the electrons in the corresponding layer 
$0\!\le\!Z\!\le\!Z_{{\scriptscriptstyle M}}$  within a cylinder
$C_r$ of radius $r\!\ge\! R/2,\, 2R/3$ respectively. By (\ref{zeta}), (\ref{blabla})  $\sigma\!=\!1.25$ gives
\  $\zeta\!\simeq\!Y_e^z(\xi_0)\!\simeq\!1.2 \!\times\! 10^{-3}$cm, \ $R\!\simeq\! 1.5 \!\times\! 10^{-3}$cm, 
\ whereas \   $\sigma\!=\!1.5$ gives \ $\zeta\!\simeq\!Y_e^z(\xi_0)\!\simeq\! 1.07 \!\times\! 10^{-3}$cm, \
$R\!\simeq\! 1.61 \!\times\! 10^{-3}$cm.
A plasma with $n_0\!\ge\! 10^{17}$cm$^{-3}$ is obtained by ionization
from an ultracold gas (typically, helium) jet in a vacuum chamber
hit by such an energetic laser pulse as soon as  the Keldysh parameters
$\Gamma_i\!\equiv\!\sqrt{U_i/{\sf k}}\!=\!\sqrt{2U_i/m{}c^2u_{e{\scriptscriptstyle}}^{{\scriptscriptstyle \perp}2}}$
for both first and second ionization become smaller than 1; here  ${\sf k}$ is kinetic energy, and  the  potentials  $U_i$ for first and second ionization are about 24eV,  54eV  respectively.
The length of the $z$-interval where both $\Gamma_i\!<\!1$ plays the role of
pulse length $l$. 
The ionization is practically complete and immediate if  $R \!\lesssim\! 13\times\! 10^{-3}$cm,
because for such field intensities the Keldysh parameter for double ionization
reaches values $\Gamma^d_i\!<\!1/100\!\ll\!1$ \cite{Puk02,JovFedTanDeNGiz12} very fast.
In \cite{Fio13} it has been shown that the model predicts no significant difference if
the modulating amplitude $\epsilon_s$ is not chosen as a gaussian (blue curve in fig. \ref{modAmp})
but rather inside the support $0\!\le\xi\!\le\!l_p$ as  the following fourth degree polynomial
\be
\epsilon_s(\xi)=\!b_p\, \left[\!\frac 14\!-\!\left(\!\frac{\xi}{l_p}\!-\!\frac 12\right)^2\! \right]^2
\theta\!\left[\!\frac 14\!-\!\left(\!\frac{\xi}{l_p}\!-\!\frac 12\right)^2\! \right]      \label{modamp}
\ee
(purple curve in fig. \ref{modAmp}), provided  $b_p, l_p$ are chosen 
so that the widths at half height of  $\epsilon_s^2$ and the pulse energy are the same;
this gives $l_p\!\simeq\!18.75\!\times\! 10^{-4}$cm, corresponding to a pulse duration
of about $6.25\!\times\! 10^{-14}$s. \
We use such polynomial modulating amplitude (\ref{modamp}) rather than the Gaussian one
to make numerical computations easier.

We adopt \ $K\!=\!2\,l_p^{-2}$ \ as maximal $K$. This, replaced in the definition of $T(\xi)$, gives
\ $T(\xi_0)\!=\!0.2$ and $T(\xi_0)\!=\!0.19$ 
if $\sigma\!\equiv\! R/\zeta\!=\!1.25$, $\sigma\!\equiv\! R/\zeta\!=\!1.5$ respectively; both
 fulfill (\ref{cond})$_2$. So we adopt \ $K\!=\!2\,l_p^{-2}$ \ as maximal $K$.
Eq.  (\ref{maxgamma})  gives \ $Z\!_{{\scriptscriptstyle M}}\!\simeq\! 0.44\,\zeta$,
$Z\!_{{\scriptscriptstyle M}}\!\simeq\! 0.36\,\zeta$; \
by  (\ref{zeta}), also condition (\ref{cond})$_1$
 is fulfilled in either case. 
The time $\bar t_1$ of inversion of the motion is very close to $t_0$ in either case.

In the central, right columns of table \ref{tab1} we summarize these and other outputs respectively for $R/\zeta\!=\!1.25,\, 1.5$.

As $n_0\!\ll \!n_c\!=\!\pi mc^2/e^2\lambda^2\!\simeq\! 1.7\!\times\! 10^{21}$cm$^{-3}$ ($n_c$ is the critical density),
we are indeed dealing with an underdense plasma. The pre-condition (\ref{cond1}) for an 
efficient slingshot loading  is fulfilled (the right-hand  side
$\pi/2\sqrt{K}\!\simeq\!21\!\times\! 10^{-4}$cm is indeed of the order of the left-hand side).
The maximal penetration $\zeta_i$ of the ions induced by the laser pulse
 is obtained from  (\ref{zeta}) replacing $m{}$ with their mass, hence $\zeta_i\!\ll\!\zeta$;
similarly, the displacement of ions due to their mutual electric repulsion is much smaller
than that of the lateral electrons (see appendix \ref{outside}).  This justifies
the description of ions as immobile.

\begin{table}
\begin{tabular}{|l|c|c|}
\hline
& $R/\zeta\!=\!1.25$ & $R/\zeta\!=\!1.5$ \\[2pt] 
\hline
pulse energy ${\cal E}$ & $\!\simeq\!5$ J &  $\!\simeq\!5$ J  \\[2pt]
wavelength $\lambda$ & $\!\simeq\!8\!\times\! 10^{-5}$cm &  $\!\simeq\!8\!\times\! 10^{-5}$cm \\[2pt]
pulse length $l_p$ & $\!\simeq\!18.75\!\times\! 10^{-4}$cm &  $\!\simeq\!18.75\!\times\! 10^{-4}$cm \\[2pt]
pulse spot radius $R$ & $\!=\!1.5 \!\times\! 10^{-3}$cm & $\!=\!1.61 \!\times\! 10^{-3}$cm \\[2pt]
$Kl_p^2$ &  =2 &  =2\\[2pt] 
initial electron density $n_0$ & $\,=\!6.4\!\times\! 10^{17}$cm$^{-3}$ & $\,=\!6.4\!\times\! 10^{17}$cm$^{-3}$ \\[2pt]
electrons' penetration $\zeta$ & $\!\simeq\!1.2 \!\times\! 10^{-3}$cm &  $\!\simeq\!1.07 \!\times\! 10^{-3}$cm \\[2pt]
time $\bar t$ of maximal penetration & $\simeq\!7.15\!\times\! 10^{-14}s$  &  $\simeq\!6.70\!\times\! 10^{-14}s$ \\[2pt]
time lapse of comeback 
$T\!_{{\scriptscriptstyle H}}/4$ & $\simeq\!5.247 \!\times\! 10^{-14}s$ & $\simeq\! 4.919 \!\times\! 10^{-14}s$ \\[2pt]
time of expulsion $t_e$ & $\simeq\!1.24 \!\times\! 10^{-13}s$ &  $\simeq\!1.162\!\times\! 10^{-13}s$ \\[2pt]
expelled layer thickness  $Z\!_{{\scriptscriptstyle M}}$ &  $\!\simeq\!0.44\,\zeta$  & $\!\simeq\!0.36\,\zeta$ \\[2pt]
expelled electrons charge $|Q|$  & $\,\ge\!1.41 \times 10^{-10}$C & $\,\ge\!1.44 \times 10^{-10}$C \\[2pt]
maximal relativistic factor $\gamma_{e{\scriptscriptstyle M}}$ & $\!\simeq\! 1.83$ & $\!\simeq\! 1.65$ \\[2pt]
maximal expulsion energy $H$ &$\!\simeq\!  0.94$ MeV & $\!\simeq\!  0.85$ MeV  \\[2pt]
maximal electric field $E^z_{{\scriptscriptstyle M}}$ & $\simeq\!1.4 \, $GV/cm & $\simeq\! 1.25 \, $GV/cm \\[2pt]
\hline
\end{tabular}
\caption {Main inputs and outputs for FLAME experiments}
 \label{tab1}  
\end {table}

\section{Conclusions and final remarks}
\label{Conclusions}

We have shown that the slingshot effect is possible if the laser pulse pancake is sufficiently short
[eq. (\ref{cond1}) and estimate after (\ref{Kxi1})], its radius is not too small [eq. (\ref{nobubble})], and the
electromagnetic field inside is sufficiently intense for the expelled electrons to have a significant
kinetic energy, i.e. a non-negligible $\gamma_{e{\scriptscriptstyle M}}\!-\!1$.
To make the latter condition quantitative and deduce conservative estimates of the
final energy of the expelled electrons, as well as of the corresponding spectrum,
we have used eq.  (\ref{maxgamma}), (\ref{gammaeM}),  (\ref{pulseEn}),  fig. \ref{gammaplot},
and assumed in addition the technical condition (\ref{cond}). 

Eq. (\ref{cond})$_1$  in particular fixes un upper bound for $K$, hence for $n_0$.
If we keep $R,\zeta$ fixed  and decrease $n_0$ then (\ref{cond})$_1$ still holds and by
(\ref{gammaeM})  $\gamma_{e{\scriptscriptstyle M}}\!-\!1$ decreases proportionally.
This scaling and the backward direction of expulsion may 
experimentally  discriminate the slingshot effect from LWF or other acceleration
mechanisms \cite{LuEtAl07}.

On the other hand, we expect that we can increase
$\gamma_{e{\scriptscriptstyle M}}$ by decreasing $R$ and increasing $n_0$
so that $\zeta$ also decreases and (\ref{nobubble})  remains fulfilled.
As this would violate the technical  \cite{Fio13} condition (\ref{cond}),
a quantitative estimate  of the corresponding slingshot effect is not possible at the 
level of approximation described here.
Preliminary numerical computations at the next level indicate that the final 
kinetic energy of expelled electrons can be thus increased by at least one order of magnitude.
This will be elaborated elsewhere.

The use of more powerful and shorter laser pulses would further increase
the final  energy of expelled electrons, while allowing the use of light solids
as targets instead of gas jets. This could be accomplished, for instance,
by the  laser  at the ELI facility, which will be able to 
shoot  linearly polarized pulses with  \ $\lambda\!\simeq\! 8\!\times\! 10^{-5}$cm, \
energy \ $\E\!\simeq\!10^{9}$erg \  and an approximately gaussian modulating amplitude
 with width at half height \ $l'\!\simeq\! 3\!\times\! 10^{-4}$cm \
\cite{Pri14b}. Preliminary estimates seem to show
that use of such a laser  would allow to further increase the final energy of expelled electrons by more
than one order of magnitude.

In the present model we have done a number of approximations. In addition to the ones 
already mentioned, we have also neglected the negative ponderomotive force of the  pulse
after its maximum has overcome the electrons, as well as
 the negative ponderomotive force of the
`reflected EM wave' generated by the impact of the pulse on the plasma;
both add to the electrostatic force to increase
the energy of the electrons in the expulsion phase. We have also
approximated the transition region from $n_e\!=\!0$ to $n_e\!=\!n_0$
as the surface $z\!=\!0$, rather than a thin layer; we expect that the latter
would modify the shape  of the spectrum in fig. \ref{gammaplot}, but not the main results.

In (\ref{pump}) we have schematized the $\rho$-dependence of the pulse by 
$\theta(R\!-\!\rho)$, rather than by a more realistic smoothly decaying  factor, such as $e^{-\rho^2/r^2(t)}$
with $r(0)\!\sim\! R$. 
Identifying $r(t)$ with $R$ is justified during the impact because  $\zeta,Z\!_{{\scriptscriptstyle M}}\!\ll\!l_R$
(the Rayleigh length  of the focalized laser pulse at FLAME is $l_R\!\sim\!2\!\times\! 10^{-2}$cm).
The rather coarse approximation of a gaussian by a step function 
is adopted to ease the computations of the finite $R$ corrections (appendix \ref{outside})
and is a further reason why our prediction 
 of the expelled charge $Q$ is an {\it underestimate} of the real one.

We have adopted all the decribed conservative estimates  to give a safe basis to the prediction
of this new effect, leaving optimization of the process as a task for further theoretical and
experimental works.

\acknowledgments
We  thank L. Gizzi for information on FLAME and valuable  suggestions,  S. De Nicola for discussions.
This work was partially supported by UniNA and Compagnia San Paolo under the grant "STAR Program 2013".

\appendix
\section{Plane plasma equations}
\label{Planesetup}

{\bf Some results of \cite{Fio13}.} \
Let $\widetilde{n_{e0}}(Z)\!\equiv\! n_{e}(0,\!Z)$ be the initial density.
In the $\widetilde{n_{e0}}\!\equiv\! 0$ limit the electron motion  
in terms of the pump \ $\bA\!^{{\scriptscriptstyle\perp}}(x) \!=\! 
\Ba^{{\scriptscriptstyle \perp}}(ct\!\!-\!z)$ \ is:
\bea
&& \bu_e^{{\scriptscriptstyle\perp}}\!(t,z)\!=\!
\frac {e\Ba^{{\scriptscriptstyle \perp}}\!(ct\!\!-\!z)}{mc^2}, \quad \:  u_e^z\!=
\frac 12\bu_e^{{\scriptscriptstyle \perp}}{}^2\!, \quad \:
  \gamma_e\!=\!
1\!+\!u_e^z,         \quad       \label{n=0'+}\\[6pt]
&& \ba{l}
z_e(t,Z)=ct-\!\Xi_e^{-1}\!\left( ct\!-\! Z \right),\\[6pt]
 Z_e(t,z)=ct\!-\!\Xi_e\!
\left( ct\!-\! z \right)=z\!-\! Y^z_e(ct\!-\! z),\\[6pt]
\bx^{{\scriptscriptstyle \perp}}_e(t,\bX )=\bX^{{\scriptscriptstyle \perp}}
\!+\!\bY^{{\scriptscriptstyle \perp}}_e\!\left[ct-\!
z_e(t,Z )\right],\\[6pt]
\bX^{{\scriptscriptstyle \perp}}_e(t,\bx)=
\bx^{{\scriptscriptstyle \perp}} \!-\!
\bY\!^{{\scriptscriptstyle \perp}}_e\!\left(ct-\!z\right),
\ea \qquad \label{hatxtxp}  \\[6pt]
&&\partial_t  Z_e =-cu_e^z, \qquad\partial_z  Z_e =\gamma_e
,\qquad\partial_Z z_e=1/ {\widetilde{\gamma_e}}.
\label{hatdtzz'}
\eea
Relations (\ref{n=0'+}) describe  in the Eulerian picture forward travelling-waves moving 
 with phase velocity equal to  $c$. In  (\ref{hatxtxp}) 
 we have used the primitives of \ $\bu_e,\gamma_e$
\be
\bY\!_e(\xi)\!\equiv\!\!\int^\xi_0\!\!\!\! dy \,\bu_e(y),
\quad\Xi_e(\xi)\!\equiv\!\!\int^\xi_0\!\!\!\!  dy\, \gamma_e(y)\!=\!
\xi  \!+\! Y^z_e\!(\xi).                                      \label{defYXi}
\ee
As $u_e^{{\scriptscriptstyle z}}\!\ge\! 0$,  \ $Y^z_e(\xi)$ \ is increasing, \
$\Xi_e(\xi)$ \ is strictly increasing (and invertible), and 
the longitudinal motion is purely forward.  
Replacing (\ref{hatdtzz'}) in the relations 
\be
\ba{l}
 n_e (t,\!z)=\widetilde{n_{e0}}[Z_e(t,\!z)]\, \partial_z  Z_e(t,\!z),  \\[6pt]
\widetilde{n_{e0}}[Z_e(t,\!z)] \,\partial_t  Z_e(t,\!z)\!+\![n_e v^z_e](t,\!z)=0.  
\ea           \qquad\qquad           \label{j_h}
\ee
one obtains the Eulerian electron density in the next approximation.
As said, here we  assume \ $\widetilde{n_{e0}}(Z)\!=\!n_0\theta(Z)$.

In Ref. \cite{Fio13} it is shown that the EM field remains close to the pump in the space-time region  
$0\!\le\!ct\!-\!z\!\le\! \xi_0$, $0\!\le\!ct\!+\!z\!\ll\!  \frac{2\pi} {K\lambda}$,
and that the motion (\ref{n=0'+}-\ref{defYXi}) is a good approximation of the real 
 forward motion of all electrons with initial positions such that \
$0\!\le\!Z\!\le\! Z_{{\scriptscriptstyle M}}$, provided  \cite{Fio13}
\be
\ba{l}
\xi_0\!+\!2Y_e^{{\scriptscriptstyle z}} (\xi_0)\!+\! 2Z_{{\scriptscriptstyle M}} \ll  \frac{2\pi} {K\lambda},
\qquad\quad T(\xi_0)\ll 1,\\[8pt]
\mbox{where}
\quad  V (\xi)\!\equiv\!\int^{\xi}_0\! dy\:Y_e^{{\scriptscriptstyle z}} (y),\\[10pt]
G(\xi)\!\equiv\!\int^{\xi}_0\! dy\,\frac{\left(1\!+\!\bu_e^2\right)
\left[e^{8KV}\!\!\!-\!1\right] }{1\!+\!\bu_e^2\!+\!e^{8KV}}\!(y),\qquad
T\!\equiv\! G/ Y_e^{{\scriptscriptstyle z}}.
\ea
\label{cond}
\ee
$T(\xi_0)$ gives the relative error between $Y_e^{{\scriptscriptstyle z}}\!(\xi_0\!)$ 
and $\zeta$ in the next approximation.

\smallskip
According to formulae (33) and (61) of \cite{Fio13}, in the next approximation
the time  when the surface electron invert their motion is
$\bar t_1\!=\!\Xi_e(\xi_1)/c$, 
where $\xi_1$
is determined by $\exp[8K V(\xi_1)]=1\!+\!\eta[e\lambda\epsilon_s(\xi_1)/2\pi mc^2]^2$
$\eta=1,1/2$ \ respectively for circular, linear polarization) as a function of the physically tunable
parameter $K$ (or, equivalently, of $n_0$).
This equation can be solved for $K$:
\be
K(\xi_1)
=\ln\big\{1\!+\!\eta\,\big[e\lambda\epsilon_s(\xi_1)/2\pi mc^2\big]^2\big\}/8 V(\xi_1)
\label{Kxi1}
\ee
[$K(\xi_1)$ is a strictly decreasing function from $K(0)\!=\! \infty$ to $K(l)\! =\! 0$]; \ 
$K[\Xi_e^{-1}(c\bar t_1)]$ will be the value of $K$ leading to the specified inversion time $t_1$.
$T(\xi)\!\ll 1\!$  for small $\xi$ and rapidly grows around $\xi_1$.
We choose $\xi_1$ so that $\xi_1\!-\!\xi_0$ is positive, as small as possible, 
but makes $K(\xi_1)$ still compatible with (\ref{cond}); this guarantees \
 $\bar t\!\simeq\!t_0\!\simeq\!\bar t_1$ \ and an efficient slingshot loading.

\begin{figure}
\includegraphics[width=8cm]{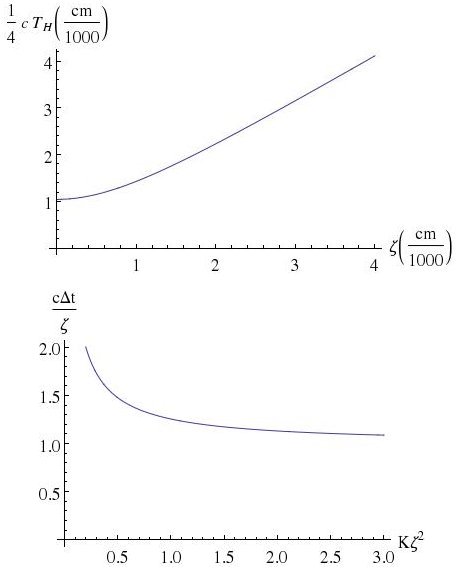}
\caption{The return time lapse $T\!_{{\scriptscriptstyle H}}/4$ for $K\!\equiv\!\pi e^2 n_0/m{} c^2\!=\!5.7\!\times\!10^{-7}\mbox{cm}^{-2}$
as a function of the penetration $\zeta$ (up), 
the ratio $cT\!_{{\scriptscriptstyle H}}/4\zeta$ as a function of $K\zeta^2$   
(down).}
\label{Deltat-Plot}
\end{figure}

\bigskip
{\bf About the relativistic harmonic oscillator.} \
Assuming for simplicity $\bu_{ei}^{{\scriptscriptstyle \perp}}\!=\!0$, 
from (\ref{conservation}) we find that  $\Delta z(t\!,Z)\!\equiv\!z_e(t\!,Z)\!-\!Z$  fulfills
\bea
&& 1+2K[\zeta^2\!-\! \Delta  z^2]=\tilde\gamma_e=1/\sqrt{1\!-\!\tilde\beta_e^{{\scriptscriptstyle z}2}}\qquad 
\Rightarrow \nn
&& \frac 1c\frac{d \Delta z} {dt}=\tilde\beta_e^{{\scriptscriptstyle z}} =-\frac {\sqrt{4K[\zeta^2\!-\!  \Delta z^2]
\!+\!4K^2[\zeta^2\!-\!  \Delta z^2]^2 }}{1+2K[\zeta^2\!-\!  \Delta z^2]}\qquad\Rightarrow\nn
&&  \Delta t=\frac 1{2c\sqrt{\!K}}\!\int^{\zeta}_{\Delta z}\! \! 
\frac{dy\,[1\!+\!2K(\zeta^2\!-\!y^2)]}{\sqrt{\!\zeta^2\!-\!y^2\!+\!K\!(\zeta^2\!-\!y^2)^2 }}   \nonumber
\eea
Therefore the  time lapse $\Delta t$ for the electrons to go 
from $z_e\!=\!\!Z\!+\!\zeta$ with zero initial velocity to  $z_e\!=\!Z\!+\!\Delta z$ is  
\be
\Delta t=\frac{\zeta}c\, f\!\left(\!\!\frac{\Delta z}{\zeta};\! K\zeta^2\!\right)\!,\quad f(D; \chi)\!\equiv\!
\!\!\int\limits^1_{D}\! \! \frac{dv[1/2\!+\!\chi(1\!-\!v^2)]}{ \sqrt{\!\chi}\sqrt{\!1\!-\!v^2\!+\!\chi(1\!-\!v^2)^2 }},
 \label{Deltat}
\ee
[$\Delta z(t\!,Z)$ \ is independent of $Z$, and so is $\Delta t$]. In particular the time needed to return
from  $z\!=\!Z\!+\!\zeta$ to their initial equilibrium position $z\!=\!Z$  is, as claimed, 
$$
 \frac 14 T\!_{{\scriptscriptstyle H}}=\frac 1{2c\sqrt{\!K}}\!\int^{\zeta}_0\! \! 
\frac{dz\,[1\!+\!2K(\zeta^2\!-\!z^2)]}{\sqrt{\!\zeta^2\!-\!z^2\!+\!K\!(\zeta^2\!-\!z^2)^2 }}.
$$
$T\!_{{\scriptscriptstyle H}}$ is the plasma oscillation period. 
In fig. \ref{Deltat-Plot}-up we plot $cT\!_{{\scriptscriptstyle H}}/4$ 
as a function of \  $\zeta$ \ for  \ $K\!=\!5.7\!\times\!10^{-7}\mbox{cm}^{-2}$ \ (a value relevant 
 for a possible experiment at FLAME, see below)  \cite{footnote5}. 
The ratio $cT\!_{{\scriptscriptstyle H}}/4\zeta$
is a function of $K\zeta^2$ only, going to 1 as  $K\zeta^2\!\to\!\infty$, see fig. \ref{Deltat-Plot}-down,
i.e. the average electron velocity goes to $c$ in this limit.
The function $f$ defined in (\ref{Deltat}) is a strictly decreasing, and therefore invertible,
 function of $D\!\in\![-1,1]$, and so is
$$
m(D; \chi) \equiv f(D; \chi)-D.
$$
Inverting (\ref{Deltat}) at fixed $K\zeta^2$ we obtain $\Delta z=f^{-1}(c\Delta t/\zeta,K\zeta^2)$. 
Setting \ $\Delta t\!\equiv\!(t\!-\!\bar t)\!-\!Z/c$ we obtain $z_e(t,\!Z)$ in the backward motion of the  electrons.
 (\ref{Deltat}) is also equivalent to $m(\Delta z/\zeta;\! K\zeta^2)\!=\!c[(t\!-\!\bar t)\!-\!z]/\zeta$;
inverting the latter we find $Z_e(t,\!z)$ in the backward motion.

\begin{figure}
\includegraphics[height=4cm]{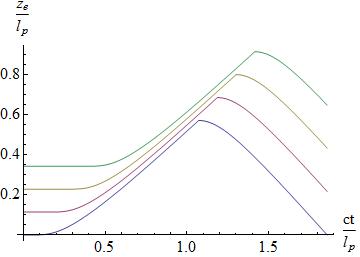} 
\caption{The graphs of the  function (\ref{completeelectronmotion}) of $t$ for the initial values $Z\!=\!0,0.2\zeta,0.4\zeta,0.6\zeta$ and other parameters (pulse length $l_p$, etc.) given in the central column of Table \ref{tab1}.}
\label{zePlot}
\end{figure}

\bigskip
{\bf The whole electron motion until expulsion.} \
Combining the last results with (\ref{hatxtxp}) we obtain our basic approximation for the whole
motion of the electrons until their expulsion  in the plane wave idealization:
\bea
z_e(t,Z)=\left\{\!\ba{ll}
ct\!-\!\Xi_e^{-1}\!\left( ct\!-\! Z \right)\quad &  t\le\bar t \!+\!Z/c,\\[6pt]
Z\!+\! \zeta f^{-1}\!\!\left\{\!\frac{c(t\!-\!\bar t)\!-\! Z}\zeta;\! K\!\zeta^2\!\right\} \:\: 
&t_e\!\ge\! t\!>\!\bar t \!+\!Z/c.\ea\right.    \qquad \label{completeelectronmotion}\\[4pt]
Z_e(t,z)=\left\{\!\ba{ll}
z\!-\! Y_e^{{\scriptscriptstyle z}}(ct\!-\! z)\quad &  ct\!-\! z\le\xi_0,\\[6pt]
z\!-\! \zeta m^{-1}\!\!\left\{\!\frac{c(t\!-\!\bar t)\!-\! z}\zeta;\! K\!\zeta^2\!\right\} \:\: & ct\!-\! z>\xi_0.\ea\right.
\qquad  \label{completeelectronmotion'}
\eea
In fig. (\ref{zePlot}) we plot the graphs of $z_e(\cdot,Z)$ (worldlines) normalized to $l_p$ 
(the pulse length available at FLAME) 
for a few fixed values of $Z$; these worldlines do not intersect, or equivalently 
the map $z\mapsto Z$ is one-to-one for each fixed $t$. 
Replacing $Z_e(t,z)$ in (\ref{j_h}) we find $n_e(t,z)$.

\begin{figure*}
\includegraphics[width=16cm]{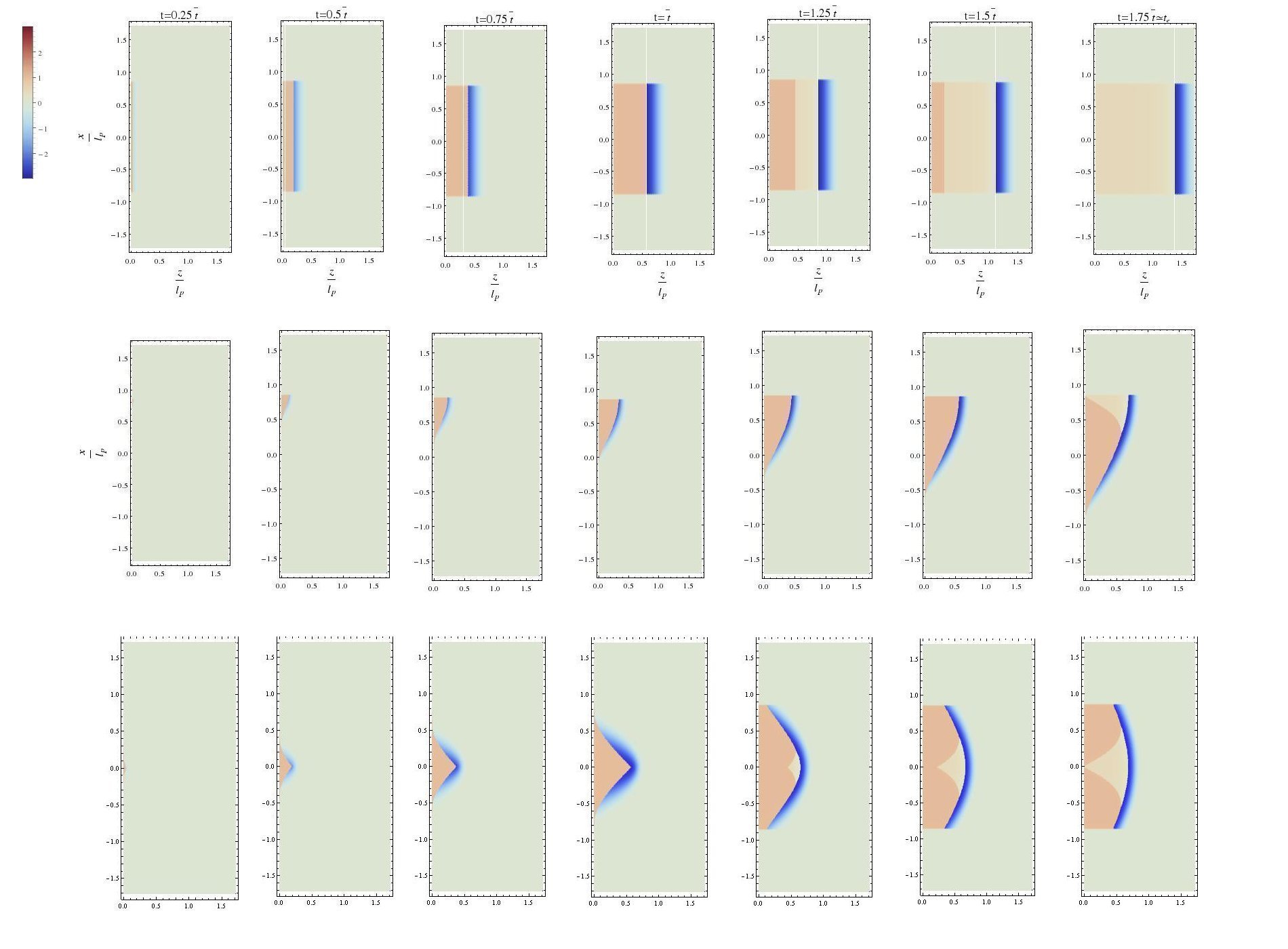}
\caption{Upper row: coloured plots  in the $z,x$ plane of  the normalized charge density 
\  $\frac{j^0(t,\!z,\!x)}{en_0}$ \  
multiplied by the characteristic function 
of the cylinder $\RC$,  at \
$t\!=\!0.25\,{\bar t},0.5\,{\bar t},0.75\, \bar t,1.25\, \bar t,1.5\,\bar t, 1.75\, \bar t$  \ 
for the possible FLAME experiment parameters $R/\zeta=1.5$, \  $K\zeta^2\!=\!0.654$; $j^0(t,\!z,\!x)$ 
is computed in the approximation  (\ref{completeelectronmotion}-\ref{completeelectronmotion'}). 
  $z,x$ are expressed in units of the length $l_p$ of the pulse. 
We name $S_0(t)$ the surface of discontinuity between the positively (brown) and 
negatively (blue) charged region. \\
Central row: corresponding coloured plots of  the retarded normalized charge density \  $\frac{j^0(t_r,\!z,\!x)}{en_0}$ \  
($t_r\!=\!t\!-\!|\bx\!-\!\bx'|/c$) at 
at the same times, as seen from the point \ $\bx\!=\!R\hat\bx$ of $\partial\RC$.  \\
Lower row: corresponding coloured plots of  the retarded normalized charge density \  $\frac{j^0(t_r,\!z,\!x)}{en_0}$ \  
($t_r\!=\!t\!-\!|\bx_e(t,\0)\!-\!\bx'|/c$) at 
the same times, as seen from the point \ $\bx_e(t,\0)\!=\!z_e(t,0)\hat\bz$ on the $\vec{z}$-axis. 
}
\label{ChargeDistributions}
\end{figure*}

\section{$R\!<\!\infty$ corrections to  electron motions}
\label{outside}

Here we study where  (\ref{completeelectronmotion}) is a good approximation of  the {\it real}
 electron motion induced by a pulse of  {\it finite}  $R$.

The general solution (retarded electromagnetic potential) of the Maxwell equation 
$\Box A^\mu\!=\!4\pi j^\mu$ in the Lorentz gauge ($\partial_\mu A^\mu\!=\!0$) 
with zero asymptotic conditions reads
\be
 A^\mu\!(t,\!\bx)\!=\!\!\int\!\! \!d^3x' \frac{j^\mu[t_r(x,\!\bx'),\bx']}{|\bx\!-\!\bx'|}, \quad       
t_r(x,\bx')\!\equiv\!t\!-\! \frac{|\bx\!-\!\bx'|}c,                     \label{retAmu}
\ee
and  $\bE(t,\!\bx)\!=\!\frac {-1}c\partial_t\bA\!-\!\!\nabla\!\! A^0\!$. 
As a first rough approximation of the real current  induced by the pump (\ref{pump})
we adopt
\be
\big(j^\mu(t,\!\bx)\big)\!\equiv\! e 
                         \Big(\widetilde{n_{e0}}(z)\!-\!n_e(t,\!z),-(n_e\bb_e)(t,\!z)\Big)\,\theta(R\!-\!\rho),
 \label{jmu}
\ee
where  \ $\widetilde{n_{e0}}(Z)\!=\!n_0\theta(Z)$  \ and  $n_e,\bb_e$ are determined from (\ref{j_h}),
(\ref{completeelectronmotion}-\ref{completeelectronmotion'});
a plot of $j^0$ is in fig. \ref{ChargeDistributions}.
As velocities are relativistic, (\ref{retAmu}) differs significantly 
from the instantaneous counterpart where $t_r$ is replaced by $t$.
It gives the EM potential (and field, after derivation) 
generated by the electric current  (\ref{jmu}), which adds to 
the pump  (\ref{pump}). We now  estimate its influence over the  electron motion 
 (\ref{jmu}) itself, and where this can be neglected.

\medskip
{\bf i) Determination of   $\widetilde{E^z}$ on the $z$-axis.} \
Below we show that if  $\widetilde{n_{e0}}(Z)$ vanishes at $Z\!<\!0$
the condition
\be
0\!\le\!z\!\le\!ct\!\le\!\sqrt{R^2\!+\!z^2}            \label{cond'}
\ee
implies
\be
 E^z(t,\!z\hat\bz)\!=\!4\pi e
\big\{\widetilde{N_{e}}(z)\!-\!\widetilde{N_{e}}[Z_e(t,\!z)]\big\},
 \label{noinfluence}
\ee
with $\widetilde{N}_e( Z )\!\equiv\!\int^{ Z }_0 \!\! dy \widetilde{ n_{e0}}(y)$  a primitive
of $\widetilde{ n_{e0}}$.
If $\widetilde{n_{e0}}(Z)\!=\!n_0\theta(Z)$  then $\widetilde{N}_e( Z )\!=\!n_0Z\theta(Z)$, and (\ref{noinfluence}) 
becomes  (\ref{elF}),
exactly as in  the $R\!=\!\infty$ case.
This should not come as a surprise: (\ref{cond'})
ensures that the spacetime region where the $R\!<\!\infty$  and the  $R\!=\!\infty$ 4-current
$j^\mu$ differ is not causally connected to  the spacetime point $(ct,z\hat\bz)$ 
\cite{footnote6}. 

Moreover, we show that $\widetilde{E^{{\scriptscriptstyle z}}}(t,\!Z\hat \bz)$
remains very close to the Lagrangian counterpart of (\ref{elF}) for a time lapse growing with $R/\zeta$,
 even if 
 the Lagrangian formulation of  (\ref{cond'}), i.e.
\ $0\!\le\!z_e(t,Z)\!\le\! ct\!\le\!\sqrt{R^2\!+\![z_e(t,Z)]^2}$, \  
is not fulfilled. Tuning $R/\zeta$
we can thus neglect their difference until the expulsion time $t_e$.
In particular  we find 
that for the suggested experiments at  FLAME
 \ $E^z(\bar t,\!\zeta\hat \bz)/4\pi en_0$ practically
coincides with 1 with both choices  $\sigma\!\equiv\! R/\zeta\!=\!1.25,\, 1.5$.

We determine $\widetilde{E^z}(t,\!Z\hat\bz)$ first  for a generic
 $\widetilde{n_{e0}}(Z)$ vanishing at $Z\!<\!0$.
Denoting the cylindrical coordinates of $\bx'$ as $(z',\rho,\varphi)$,
from (\ref{retAmu}-\ref{jmu})  \ it follows
\bea
 &&
\ba{l}\frac 1{2\pi e} E^z(t,\!z\hat \bz)=-\frac 1{2\pi e}\left[\frac 1c\partial_tA^z\!+\!\partial_zA^0\right]
\ea \nn
&& \!=\!\!\int\limits^{2\pi}_0\!\! \frac {d\varphi}{2\pi}\!\!\int\limits^{ct}_0\!\!
 dz'\!\!\!\int\limits^R_0\!\!\rho d\rho \! \left\{\!\frac{\partial_t(n_e\beta_e^z)(t_r,\!z')} {c\sqrt{\!(z\!-\!z')^2\!+\!\rho^2}}-
\partial_z \frac{\widetilde{n_{e0}}(z')\!-\!n_e(t_r,\!z')}{\sqrt{\!(z\!-\!z')^2\!+\!\rho^2}}\! \right\} \nn
&&\!=\!\!\int\limits^{ct}_0\!\! \! dz'\!\!\!\int\limits^R_0\!\!\! d\rho \,
\partial_\rho\!\left\{\!\frac{(z\!-\!z') [n_e(t_r,\!z')\!-\!\widetilde{n_{e0}}(z')]}{\sqrt{(z\!-\!z')^2\!+\!\rho^2}}-(n_e\beta_e^z)(t_r,\!z') \!\right\}\nn
  &&\!=\!\!\int\limits^{ct}_0\! \! dz'\!
\left\{\!\frac{(z\!-\!z')[n_e(t_r,z')\!-\!\widetilde{n_{e0}}(z')]}{\sqrt{(z\!-\!z')^2\!+\!\rho^2}}
 -(n_e\beta_e^z)(t_r,z')\!\right\}_{\!\!\rho\!=\!0}^{\!\!\rho\!=\!R}     \nonumber
\eea
\bea
&&\!=\!\!\int\limits^{ct}_0\! \! 
dz' \!\left\{\!\frac{(z\!-\!z')\left\{[\widetilde{n_{e0}}(Z_e)
\partial_y Z_e](ct_r,y)|_{y\!=\! z'}\!-\!\widetilde{n_{e0}}(z')\right\}}{\sqrt{(z\!-\!z')^2\!+\!\rho^2}}\right.\nn
&& \left.\qquad +\frac 1c[\widetilde{n_{e0}}(Z_e)\partial_t Z_e](t_r,\!z')\!\right\}_{\rho\!=\!0}^{\rho\!=\!R};
 \label{Ezgen}
\eea
in the derivation we have used: 
$ct_r\!=\!ct\!-\!\sqrt{(z\!-\!z')^2\!+\!\rho^2}$ for $\bx\!=\!z\hat \bz$;
for any differentiable function $f(s)$ the identities 
$\rho\partial_z f\big[t_r\big]\!=\!\partial_\rho (z\!-\!z')f(t_r)$,  
$\frac{-\rho}{c\sqrt{(z\!-\!z')^2\!+\!\rho^2}}\partial_t f(t_r)\!=\!\partial_\rho
f\big(t_r)$; \ (\ref{j_h})
in the last equality.
If  $\widetilde{n_{e0}}(Z)\!=\!n_0\theta(Z)$  we find
\bea{}
&&\frac{E^z(t,\!z\hat \bz)}{2\pi en_0}\!=\!\sqrt{\!(z\!-\!ct)^2\!+\!R^2}\!-\!
\sqrt{\!z^2\!+\!R^2}\!-\!|z\!-\!ct|\!+\!|z|   \label{Ez}
\\ && \!+\!\!\int\limits^{ct}_0\! \! 
dz' \!\left\{\!\left.\frac{(z\!-\!z')[\theta(Z_e)\partial_y Z_e](t_r,\!y)}{\sqrt{(z\!-\!z')^2\!+\!\rho^2}}\right|_{y\!=\! z'}\!\!\!\!\!\! +\!\frac{\theta(Z_e)\partial_t  Z_e}c(t_r,\!z')\!\right\}_{\!\!\rho\!=\!0}^{\!\!\rho\!=\!R}.
\nonumber
\eea
For the suggested FLAME experiment
the numerical evaluation by {\it Mathematica} in the  approximation (\ref{n=0'+}-\ref{defYXi}), (\ref{completeelectronmotion}-\ref{completeelectronmotion'})
gives \  $E^z(\bar t,\!\zeta\hat \bz)/4\pi en_0\!=\!.999$, 
$E^z(\bar t,\!\zeta\hat \bz)/4\pi en_0\!=\!1$,  \
respectively for $\sigma\!\equiv\! R/\zeta\!=\!1.25$,  $\sigma\!\equiv\! R/\zeta\!=\!1.5$.
as claimed after equation (\ref{noinfluence}).

\medskip
{\bf Proof of (\ref{noinfluence}).} \
In (\ref{Ezgen}) it is \ $t_r\!=\!t\!-\!\sqrt{R^2\!+\!(z\!-\!z')^2}$ 
\ for $\rho\!=\!R$; from (\ref{cond'}) and $z'\!\ge\!0$ it follows \ $ct_r\!\le\!z'$ 
\cite{footnote7}, whence \ $z_e(t_r,Z)\!=\!Z$,
 $Z_e(t_r,z')\!=\!z'$, $\partial_{z'}Z_e(t_r,z')\!=\!1$, $\partial_{t}Z_e(t_r,z')\!=\!0$, 
so that the integrand  in (\ref{Ezgen}) vanishes for $\rho\!=\!R$.  \  
For $\rho\!=\!0$ it is \ $t_r\!=\!t\!-\!|z'\!\!-\!z|/c$, \ and  (\ref{Ezgen}) gives 
\bea{}
 &&\frac{E^z\!(t,\!z\hat \bz)}{2\pi e} \!=\! \int^{ct}_0\! \!\! dz' \Big[s(z,\!z')
\{[\widetilde{n_{e0}}(Z_e)\partial_y Z_e]\left(t\!-\!|z'\!\!-\!z|\!/c,\!y\right)\!|_{y\!=\! z'}  \nn
&& \qquad  - \widetilde{n_{e0}}(z')\} \!-\!\frac 1c[\widetilde{n_{e0}}(Z_e)
\partial_t Z_e](t\!-\!|z'\!\!-\!z|/c\,,z')\Big]\nn
&&\!=\!\int^{ct}_0\! \! dz' \, s(z,\!z') 
\Big\{\widetilde{n_{e0}}(Z_e) \partial_{z'}[Z_e(t\!-\!|z'\!\!-\!z|/c\,,z')]\!-\!\widetilde{n_{e0}}(z')\!\Big\}\nn
&&\!=\!\int^{ct}_0\!\! \! dz' \, s(z,\!z') 
\Big\{ \partial_{z'}\widetilde{N_{e}}[Z_e(t\!\!-\!|z'\!\!-\!z|/c\,,z')]\!-\!\widetilde{n_{e0}}(z')\Big\}\nn
&&\!=\!\int^{ct}_z\! \! dz' \Big\{\partial_{z'}\widetilde{N_{e}} [Z_e(t\!\!+\!(z\!-\!z')/c\,,z')]\!-\!n_{e0}(z')\Big\}
 \nn&& \qquad -\!\!\int^z_0\! \!\! dz' \Big\{\partial_{z'}\widetilde{N_{e}} [Z_e(t\!\!+\!(z'\!\!-\!z)/c\,,z')]\!-\!
\widetilde{n_{e0}}(z')\Big\} \nn
&& \!=\! \widetilde{N\!_{e}}\!\left[Z_e\!\left(\frac zc,\!ct\right)\right]\!-\!\widetilde{N\!_{e}}\!(ct)\!+\!2\Big\{\!\widetilde{N\!_{e}}(z)\!-\!\widetilde{N_{e}}[Z_e(t,\!z)]\!\Big\}\nn
&& \!+\!\widetilde{N\!_{e}}\!\!\left[Z_e\!\left(\!t\!-\!\frac zc,\!0\!\right)\!\right]\!=\!  2\Big\{\widetilde{N\!_{e}}(z)\!-\!\widetilde{N\!_{e}}\!\left[\!Z_e\left(\!t,\!z\!\right)\right]\!\Big\}
\eea
where $s(z,\!z')$ stands for the sign of $z'\!\!-\!z$;
in the last equality we have used $Z_e(z/c,\!ct)\!=\!ct\!>\!0$, $Z_e(t,z)\!>\!0$,
$Z_e(t\!-\!z/c,0)\!<\!0$,  $\widetilde{N_{e}}[Z_e(t\!-\!z/c,0)]\!=\!0$. 
Note that the condition $ct_r\!\le\!z'$  is  equivalent to requiring that $t$ is less than the sum of the time
necessary for the pulse to reach any point $\bx'\!\in\!\partial\RC$ and of the time lapse
necessary for a EM signal to travel from  $\bx'$ to $\bx\!=\!z\hat\bz$, what ensures 
by causality also that at such $t$ the fields in $\bx$ have
not been influenced by the plasma  outside $\partial\RC$, as anticipated.

\bigskip
{\bf ii) Upper bound for $\widetilde{E^\rho}$  and for 
the inner displacement of the electrons outside $\RC$.} \
As our goal here is to find a sufficient condition for a significant expulsion (expulsion for all electrons initially located in $C_r$,
 $r\!\gtrsim\!R/2$, say), for simplicity of computations we overestimate the $\sigma$ of eq. (\ref{nobubble}) as follows.
For all electrons initially located at points $\bX\!\simeq\!R\hat\brho\!+\!Z\hat\bz$ \ (with some unit vector $\hat\brho\!\equiv\!\cos\Phi\hat\bx\!+\!\sin\Phi\hat\by$) \ around the lateral boundary $\partial\RC$ of $\RC$ we:
a) neglect the initial outward boost induced by the pulse on them; b)  overestimate the inward component $e\widetilde{E^\rho}(t,\!\bX)\!\equiv\!eE^\rho[t,\!\bx_e(\bX)]$ of the electric force exerted on them by the charge distribution within $\RC$, and consequently their inward displacement $|\Delta \rho_e|$, by  the upper bound \ $\!E^\rho(t,\!R\hat\brho)$ 
\cite{footnote8}; 
c)  overestimate $1/\widetilde{\gamma_e}\!=\!1/\sqrt{1\!+\!\widetilde{\bu_e}^2}$ by $1/\sqrt{1\!+\!\widetilde{u^x_e}^2}$
in their relativistic equation of  motion. 
We show that for $t$ in a time lapse growing with $\sigma$
\be
\bE(t,\!\bx)\!\simeq\!E^\rho\!(t,\!\bx)\hat\brho\!+\!E^z\!(t,\!\bx)\hat\bz,\quad
E^\rho(t,\!\bx)\!\ll\! \widetilde{E^z}(t,\0)           \label{latE}
\ee 
on all points $\bx\!\in\!\partial \RC$, i.e. there the electric field 
 is essentially in the $\rho$-direction and much less than the longitudinal 
one experienced by the $\bX\!\!=\!\!\0$ electrons. 
This is due to geometrical reasons (a look at the coloured charge density plot of 
fig. \ref{ChargeDistributions} center and down may help in getting a qualitative understanding): 
on one hand, to the delay inherent to the retarded potential itself; 
on the other, to the fact that the contributions generated by the ions and by the forward
boosted electrons sum up on their surface $S_0(t)$ of separation, 
while they partially cancel on $\partial \RC$.
Therefore tuning $\sigma$ we can make $|\Delta \rho_e(t_e,\!R\hat\brho\!+\!z\hat\bz)|\!\le\! R\!-\!r$
for all $z$, i.e. the inward displacement  of all surface electrons at the estimated time of expulsion
$t_e \!=\!\bar t\!+\!\Delta t$ small enough not to obstruct the way out to the electrons within $C_r$.
In particular we find that for the suggested experiments at FLAME
it is sufficient to choose  $\sigma\!\equiv\! R/\zeta\!=\!1.25,\, 1.5$ \ to respectively obtain $r/R\!\ge\! 1/2,\, 2/3$.

For circular polarization 
$E^\rho$,  $\widetilde{E^\rho}$ are strictly the same on all the points of this circle
$\rho\!=\!R$,  $z\!=\!0$, in particular on the point $\bx \!=\! R \hat \bx$:  \
$E^\rho(t,\! R \hat \bx)\!=\! E^x(t,\! R \hat \bx)$. \
For other (in particular linear) polarizations the relative variation of $E^\rho$, 
$\widetilde{E^\rho}$ along the circle is negligible for modulated 
periodic oscillations, since the charge distribution has almost cylindrical
symmetry due to the fact that transverse displacements almost average to
zero in each cycle; therefore also in this case \
$E^\rho(t,\! R \hat \bx)\!\simeq\!E^x(t,\! R \hat \bx)$. \
By (\ref{retAmu}-\ref{jmu})
\bea
&& \qquad E^x(t,\! x \hat \bx)=-\frac 1c\partial_t A^x\!-\!\partial_xA^0 \nn
&& \!=\! e\! \int^{2\pi}_0\! \! \!\! \!\! 
d\varphi\!\int^{ct}_0\!\!\!\!  dz\!\int^R_0\!\!\!\! \rho d\rho 
\left\{\frac{\partial_t\,(n_e\beta_e^x)(t_r,z)} {c\sqrt{z{}^2\!+\!C}}-
\partial_x \frac{n_0\!-\!n_e(t_r,z)}{\sqrt{z{}^2\!+\!C}} \right\}  \nn 
&& \!=\! e\! \int^{2\pi}_0\! \! d\varphi\!\int^{ct}_0\! \! dz
\!\int^R_0\!\!\rho d\rho \left\{\frac{\partial_t(n_e\beta_e^x)(t_r,z)
\!+\!c\partial_xn_e(t_r,z)} {c\sqrt{z{}^2\!+\!C}} \right.
 \nn &&  \qquad \!+\!\left.
\frac{
[n_0\!-\!n_e(t_r,z)](x\!-\!\rho\cos\varphi)}{
|z{}^2\!+\!C|^{3/2}
} 
\right\}  
\label{Ex}
\eea
where  $C\!=\!x^2\!+\!\rho^2\!-\!2x\rho \cos\varphi$, \
 $ct_r=\!ct\!-\!\sqrt{z{}^2\!+\!C}$ for $\bx\!=\!x\hat \bx$.

Replacing  the  approximated  $Z_e$ (\ref{completeelectronmotion'}) 
in  (\ref{j_h}) and the resulting densities in eq. (\ref{Ex}), 
integrating the relativistic equations of motion of the electron
\ $-e\widetilde{E^x}\!=\!d\widetilde{p_e^x}/dt\!=\!mc d\widetilde{u_e^x}/dt$,  $dx_e/dt \!=\!
c\widetilde{\beta_e^x}\!=\!
c\widetilde{u_e^x}/\widetilde{\gamma_e}$ with
initial conditions \ $\widetilde{p_e^x}(0,\!x \hat x)\!=\!0$, \  
 \ $x_e(0,\!x \hat x)\!=\!x \hat x$ ($x\!\ge\! R$), \ 
using overestimates b), c) and {\it Mathematica} we numerically find 
the bounds of table \ref{tab2} for \ $\widetilde{\beta_e^\rho}(\bar t,\!R \hat \brho)$,
  $\Delta \rho_e(t,\!R\hat\brho)\!=\!\Delta x_e(t,\!R\hat\bx)$ \
in the suggested FLAME experiments.

\begin {table}
 \begin{tabular}{|l|c|c|}
\hline 
& $R/\zeta\!=\!1.25$ & $R/\zeta\!=\!1.5$ \\[2pt] 
\hline 
$|\widetilde{\beta_e^\rho}(\bar t,\!R \hat \brho)|$ &  $\,<\!0.154$ & $\,<\!0.136$ \\[2pt] 
$|\widetilde{\Delta\rho_e}(\bar t,\!R \hat \brho)|/R$ & $\,<\!0.059$ & $\,<\!0.04$ \\[2pt]
average $\tilde \beta_e^\rho$ in $[0,\bar t]$   &  $\,\simeq\!0.041$  & $\,\simeq\!0.03$ \\[2pt]
$|\widetilde{\beta_e^\rho}(t_e,\!R \hat \brho)|$ & $\,<\!0.7$ & $\,<\!0.58$ \\[2pt]
$|\widetilde{\Delta\rho_e}(t_e,\!R \hat \brho)|/R$ & $\,<\!0.50$ & $\,<\!0.33$\\[2pt]
average $\tilde \beta_e^\rho$  in $[\bar t,t_e]$ & $\,<\!0.43$ &  $\,<0.32$ \\[2pt]
\hline
\end{tabular}
\caption {Main outputs of the inward motion of the lateral electrons
in the proposed  FLAME experiments}
 \label{tab2}  
\end {table}

\medskip
Finally, we argue that if in (\ref{pump}) we replace $\theta(R\!-\!\rho)$ by $e^{-\rho^2/R^2}$, then
outer electrons cause a smaller obstruction to 
the expulsion of the $\bX\!\sim\!\0$ electrons. In fact, also the outer  electrons with initial $\rho\!\ge\! R$  
get some displacement $\Delta z\!>\!0$, therefore (attracted by the ions) cover a smaller inward distance
$\Delta \rho_e$ in the time necessary for the $\bX\!\sim\!\0$ electrons to come back  to $\bx\!\sim\!(0,0,\Delta z)$,
and therefore cannot intercept those in a larger cylinder $C_r$.


\begin{thebibliography}{0}

\bibitem{Tajima-Dawson1979} T. Tajima, J.M. Dawson, \emph{Phys. Rev. Lett.} \textbf{43} (1979), 267.

\bibitem{Gorbunov-Kirsanov1987} L.M. Gorbunov, and  V.I. Kirsanov, \emph{Sov. Phys. JETP} \textbf{66}, 290 (1987).

\bibitem{Sprangle1988} P. Sprangle, E. Esarey, A. Ting, and G. Joyce, \emph{Appl. Phys. Lett.} \textbf{53}, 2146 (1988).

\bibitem{Irman2007} A. Irman, M.J.H. Luttikhof, A.G. Khachatryan, F.A. van Goor, J.W.J. Verschuur, H.M.J. Bastiaens,  K.-J. Boller, \emph{J. Appl. Phys.} \textbf{102} (2007), 024513.

\bibitem{Joshi2006}C. Joshi, 
Scientific American \textbf{294} (2006), 40.


\bibitem{FauEtAl04}
J. Faure, Y. Glinec, A. Pukhov, S. Kiselev, S. Gordienko, E. Lefebvre, J.-P. Rousseau, F. Burgy, V. Malka, 
 Lett. Nat. {\bf 431}, 541–544 (2004).

\bibitem{Kalmykov2009} S. Kalmykov, S. A. Yi, V. Khudik, and G. Shvets, \emph{Phys. Rev. Lett.} \textbf{103} (2009), 135004.

\bibitem{Wang2013} X. Wang, {\it et al.},
\textit{Nature Communications} \textbf{4} (2013), 1988.

\bibitem{Fio13} G. Fiore,
J. Phys. A: Math. Theor. \textbf{47} (2014), 225501.

\bibitem{Fio14} G. Fiore,
Acta Appl. Math. \textbf{132}   (2014),  261-271. 


\bibitem{GizEtAl13}
L.A. Gizzi, {\it et al.},
\textit{Appl. Sci.}, {\bf 3} (2013), 559-580; doi:10.3390/app3030559.

\bibitem{Pri14b}
L. Pribyl, L. Juha, G. Korn, et al., in Proc. IBIC2012,
Tsukuba, Japan (2012), pp. 482–485.


\bibitem{Eremin2010} V.I. Eremin, A.V. Korzhimanov, A.V. Kim, \textit{Phys. Plasmas} \textbf{17} (2010), 043102.

\bibitem{Geindre2010} J.P. Geindre, R.S. Marjoribanks, P. Audebert,  \textit{Phys. Rev. Lett.} \textbf{104} (2010), 135001.

\bibitem{Gonoskov2011} A.A. Gonoskov,  A.V. Korzhimanov,  A.V. Kim,  M. Marklund,  A.M. Sergeev,
\textit{Phys. Rev.} \textbf{E84} (2011), 046403.


\bibitem{footnote1}
$\tau\!=\!\tau_1\!+\!\tau_2$, with
$\tau_1,\tau_2$ the time lapses spent during the oscillation respectively in the domain of 
the elastic and of the constant force; due to the size of the domains it is 
$\tau_1\!\ge\!T\!_{{\scriptscriptstyle H}}^{{\scriptscriptstyle nr}}/2$. 
Hence $\tau$ depends on $Z$ and on the oscillation amplitude, but is at the least of the 
order of $T\!_{{\scriptscriptstyle H}}^{{\scriptscriptstyle nr}}$.




\bibitem{footnote2}
The longitudinal eq. of motion of the $Z\!=\!0$ electrons is
$$
\frac{d\widetilde{p^{{\scriptscriptstyle z}}}}{dt}=m{}
[F_{{\scriptscriptstyle p}}^{{\scriptscriptstyle z}}
\!(ct\!-\!z_e)-\omega^2_pz_e],\quad
F_{{\scriptscriptstyle p}}^{{\scriptscriptstyle
z}}\!(\xi)\!\equiv\!\frac{(e\lambda)^2\eta\:\partial_\xi\epsilon_s^2(\xi)}{8(m{}\pi)^2c^3\gamma(\xi)},
$$
where $\eta=1,1/2$ \ respectively for circular, linear polarization, and
$\omega_p^2\!=\!4\pi e^2n_0/m{}$ is the square of the plasma frequency.
In the  nonrelativistic regime we find $\tilde\gamma\!\simeq\!1$, $z_e(t)\!\ll\! ct$,
$\tilde p^z\!\simeq\! m{}\dot z_e$, hence the maximum of  $\epsilon_s^2(\xi)$ and
the end of the pulse reach these electrons respectively at \ $t\!\simeq\!\bar t\!=\!\xi_0/c$, 
$t\!\simeq\!l/c$, \ and the equation of motion in the bulk
reduces to that of a forced harmonic oscillator: 
$\ddot z_e=F_p^z(ct)\!-\!\omega^2_pz_e$;
the solution with initial conditions $z_e(0)\!=\!\dot z_e(0)\!=\!0$ is
$z_e(t)\!=\!\int^t_0\! dt'  F_p^z(ct')\sin[\omega_p(t\!-\!t')]/\omega_p$.
Electrons initially located at small $Z\!>\!0$ 
will experience approximately the same longitudinal displacement with a time delay $Z/c$.
 \ $\tilde F_p^z \tilde v^z\!=\! F_p^z\dot z_e$ \ keeps
nonnegative if during the pulse $\dot z_e(t)\!=\!\int^t_0\! dt'  F_p^z(ct')\cos[\omega_p(t\!-\!t')]$
switches from positive to negative at $t\!=\!\bar t\!=\!\xi_0/c$ (only).
In particular it must be $z_e(t)\!\ge\! 0$ for all $t\!\le l/c$, 
 and $\xi_0,n_0,l$ must fulfill
$\pi/4\!<\!\xi_0\omega_p/c\!<\!3\pi/4 $.

If the pulse is so intense to make electrons relativistic the conditions for maximal
slingshot loading are more complicated;
for an estimate in the proposed experimental conditions see the comments after (\ref{Kxi1}).





\bibitem{footnote3}
In (\ref{elPE}) we have chosen the additive constant so that the minima of $U$
w.r.t. $z_e$ are equal to zero for all $Z$.



\bibitem{footnote4}
To perform the work integral we have used the relation \ $\sqrt{ z^2\!+\!R^2}
=\frac 12 \partial_z\!\left[z\sqrt{ z^2\!+\!R^2}\!+\!R^2\sinh^{-1}\!\frac{z}R\right]$. \
We have chosen  the additive constant equal to the last
three terms so that \ $\rU(0^-,\!Z)\!=\!U(0^+,\!Z)$.

\bibitem{JovFedTanDeNGiz12}
D. Jovanovi\'c, R. Fedele, F. Tanjia, S. De Nicola,  L. A. Gizzi,
\textit{Eur. Phys. J.} {\bf  D66} (2012), 328.

\bibitem{Puk02}
A. Pukhov,
\textit{Rep. Prog. Phys.} {\bf  65} (2002), R1-R55.

\bibitem{LuEtAl07}
W. Lu,  M. Tzoufras, C. Joshi, F. S. Tsung, J. Vieira, R. A. Fonseca,  L. O. Silva,
Phys. Rev. ST Accel. Beams  {\bf  10} (2007), 061301.


\bibitem{footnote5}
As $\partial T\!_{{\scriptscriptstyle H}}/\partial\zeta|_{\zeta=0}\!=\!0$, 
only in the nonrelativistic limit
$T\!_{{\scriptscriptstyle H}}\!\stackrel{\zeta\!\to\!0}{\simeq}\!T\!_{{\scriptscriptstyle H}}^{{\,\scriptscriptstyle nr}}\!=\!
\pi/c\sqrt{K}\!=\!\sqrt{\pi m{}/e^2n_0}$ is independent of $\zeta$
(nonrelativistic harmonic oscillator).




\bibitem{footnote6}
In fact, (\ref{cond'}) implies   $t\!<\!(z'\!+\!|\bx-\bx'|)/c$ for all  $\bx'\!\notin\!\RC$;
$z'/c$ is the time when the  $R\!=\!\infty$  pulse reaches $\bx'$,  $|\bx-\bx'|/c$ is  the time lapse necessary 
for a EM signal to travel from  $\bx'$ to $\bx$, hence their sum is  the time when 
the information that the charge distribution does not extend indefinitely in the $xy$ plane, but
is confined in $\RC$, arrives  at $\bx$.




\bibitem{footnote7}
In fact,  by the assumptions we find $-\!2ctz'\!\le\!-\!2zz'$,  
$(ct\!-\!z')^2\!=\!(ct)^2\!-\!2ctz'\!+\! z'{}^2\!\le\!R^2\!+\!z^2-\!2zz'\!\!+\! 
z'{}^2\!=\!\!R^2\!+\!(z\!-\!z')^2$ and, taking the square root,  $ct_r\!\le\!z'$.




\bibitem{footnote8}
In fact, $e\tilde E^\rho$ 
decreases as $Z$ grows and when the electrons penetrate inside  $\RC$, as correspondingly
 the negative contribution due to the  electrons displaced
beyond $S_0(t)$ grows, while the positive one by the ions decreases.







\end{thebibliography}
\end{document}